%% file: main.tex
\documentclass[11pt]{article}
\usepackage[a4paper,margin=2.4cm]{geometry}
\usepackage[T1]{fontenc}
\usepackage{lmodern}
\usepackage{amsmath,amssymb}
\usepackage{booktabs}
\usepackage{placeins}
\usepackage{float}
\usepackage{graphicx}
\graphicspath{{figures/}}
\usepackage{tikz}
\usetikzlibrary{positioning,calc,decorations.pathreplacing}
\usepackage{caption}
\usepackage{needspace}
\definecolor{decc}{HTML}{B5651D}\definecolor{decf}{HTML}{FBEBDD}
\definecolor{wordc}{HTML}{1F4E8C}\definecolor{wordf}{HTML}{E3ECF7}
\definecolor{addc}{HTML}{555555}\definecolor{addf}{HTML}{F2F2F2}
\providecommand{\ket}[1]{\lvert #1\rangle}
\usepackage{hyperref}

\usepackage{xcolor}
\usepackage{amsthm}
\newtheorem{proposition}{Proposition}
\hypersetup{colorlinks=true,linkcolor=blue!50!black,citecolor=blue!50!black,urlcolor=blue!50!black}
\definecolor{paperlink}{HTML}{21566C}
\hypersetup{linkcolor=paperlink,citecolor=paperlink,urlcolor=paperlink}
\input{numbers}

\input{frontier_numbers}


\newcommand{\phaseMaxChange}{2.45}

\title{Shared Lookups and Lazy Residues:\\ Fewer Toffoli Gates for Factoring RSA-2048}
\author{Monit Sharma\thanks{School of Computing and Information Systems,
Singapore Management University, Singapore.}
\and Hoong Chuin Lau\footnotemark[1]\ \thanks{Corresponding author:
\href{mailto:hclau@smu.edu.sg}{\nolinkurl{hclau@smu.edu.sg}}.}}
\date{}

\begin{document}
\maketitle

\begin{abstract}
In residue-number-system circuits for factoring, thousands of small primes
read the exact same quantum exponent. We exploit this shared address to
return table entries for many primes in a single lookup. Within each prime,
retaining intermediate residues lets us uncompute multiplications by simple
subtraction rather than inverse modular arithmetic, which in turn eliminates
per-step modular reductions. Combining these ideas with compressed logarithm
registers and a final multiplication fused directly into the
Chinese-remainder accumulator yields substantial gate savings across the
qubit budget. For RSA-2048 with repetition parameter $s=8$, our circuits
require an estimated $2.99$ billion Toffolis at 1,450 logical qubits and
$1.35$ billion at 2,050—24.0\,\% and 54.4\,\% below a serial arithmetic
model retuned at the same qubit caps. Relaxing $s$ to $3$ projects $0.828$ billion Toffolis at 2,998 qubits under the nominal run count and keep-rate model.
These estimates come from eight
sampled evaluations per circuit together with an overflow allowance
extrapolated from sampled primes. On Gidney's surface-code model, the
1,450-qubit circuit fits under one million physical qubits and factors in
2.7 days, whereas a 2,925-qubit circuit finishes in 16 hours on 3.2 million
physical qubits—illustrating that minimizing Toffoli count does not
automatically minimize physical runtime.
\end{abstract}

\section{Introduction}

Shor's factoring algorithm reduces integer factorization to finding the
multiplicative order of a base modulo the integer to be
factored~\cite{shor1997}, and most of the quantum circuit's cost lies in
evaluating that power coherently. A direct implementation carries out
long-register modular multiplications one at a time, whereas a residue number
system (RNS) trades register width for many short calculations modulo small,
known primes that can reuse the same workspace. The price for that space
saving is that the circuit must repeatedly consult lookup tables, build
intermediate residues, and erase them before moving on to the next prime.

This trade has practical consequences for resource estimates. Gidney and
Eker{\aa} estimated that a surface-code computer with about 20~million
physical qubits could factor RSA-2048 in eight hours~\cite{gidney2021}, and
Gidney later showed—drawing on the approximate residue arithmetic of
Chevignard, Fouque and Schrottenloher~\cite{cfs2024}—that an RNS circuit
fits below one million physical qubits with a runtime under a
week~\cite{gidney2025}. In that construction, the exponent register occupies
most of the logical memory, leaving the remaining qubits to hold the active
prime's residue, the running accumulator, and arithmetic scratch. How those
remaining qubits are allocated governs both the total gate count and the time
spent waiting for lookups.

At Gidney's published settings, roughly 61\,\% of the benchmark Toffolis go
to loop~1 (building the discrete logarithms) and 23\,\% to cleaning up
loop~3 (reconstructing residues). Two observations attack both bottlenecks
at once. First, although each prime needs a different table entry, every
prime reads the \emph{same address}—a short window of the exponent
register—so a single lookup can return a packed vector of entries for $k$
primes, paying the address-decoding cost once while keeping separate running
logarithms. Second, modular multiplication is expensive to undo only because
it erases its input; retaining that input until cleanup lets the circuit
uncompute by subtracting the exact terms it produced, which in turn removes
the need to reduce intermediate residues back into their canonical range.
Because these two ideas require extra qubits at different points in the
computation, the overall peak depends sensitively on when registers are
allocated and released.

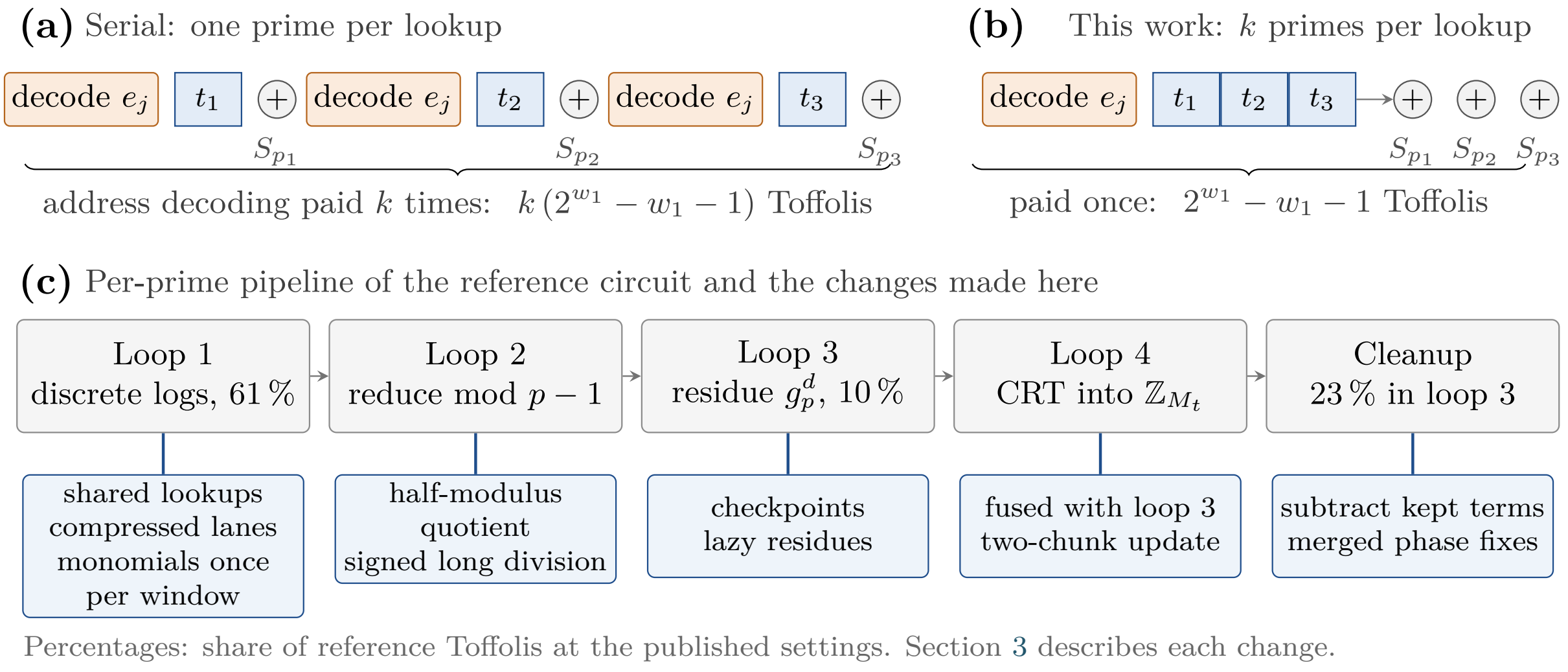
\begin{figure}[t]
\centering
\resizebox{\linewidth}{!}{\input{tikz_overview}}
\caption{The two observations behind this work. (a) In a serial residue
circuit every prime decodes the same exponent window $e_j$ before reading
its own table word $t_p$. (b) Concatenating the words of $k$ primes into one
table row pays for the address decoding once; each prime keeps its own
running sum $S_p$. (c) Where the changes act in the per-prime pipeline.
Retaining the input $x$ of a multiplication (checkpoints) lets the cleanup
subtract the terms it produced, which in turn lets the residues stay
unreduced (lazy).}
\label{fig:overview}
\end{figure}

Li et al.~\cite{li2026} study the same family of factoring circuits using a
compiler guided by language-model search, reporting sub-billion-Toffoli
circuits with selected operating points verified across all constituent
primes. We draw on their techniques for lookup synthesis, compressed
logarithms, division and phase cleanup (summarised in
Fig.~\ref{fig:overview}), and ask a sharper question about sharing: at a
fixed peak logical-qubit budget, how much can multi-prime lookup sharing
save over a serial ($k=1$) implementation that is equally free to retune its
arithmetic and parameters? We carry this comparison through to a circuit
that preserves the benchmark's truncated-modulus mask, and discuss the
differences in counting and verification conventions in Sec.~\ref{sec:related}.

\paragraph{The fixed-budget comparison.}
Let $Q$ denote the largest number of logical qubits live at the same time.
For each cap $Q$, we compare the selected circuit against the least expensive
\emph{modelled} serial setting that fits under that cap. The serial reference
processes one prime at a time ($k=1$). It can vary the repetition parameter
$s$, prime size, window widths and accumulator precision across Gidney's
\frontierGrid{}-setting grid. It also receives the applicable refinements
listed in Appendix~\ref{app:bench}. If $\Theta_1$ denotes this serial
grid, the reference cost and fractional saving of a circuit $C$ are
\[
 T_1(Q)=\min_{\theta\in\Theta_1:\,Q(\theta)\le Q}T_{\rm fact}(\theta),
 \qquad
 \Delta(C;Q)=1-\frac{T_{\rm fact}(C)}{T_1(Q)} .
\]
We evaluate the saving at the measured peak of $C$. A serial setting can
use fewer qubits or a different number of executions. When the feasible
set is empty, there is no fixed-budget comparison.

\begin{table}[htbp]
\centering\small
% Generated by audit/build_paper_assets.py
\setlength{\tabcolsep}{4.5pt} \begin{tabular}{rrrrrrrr}
\toprule
$Q$ & $s$ & $k$ & \shortstack{$T_{\rm shot}$\\$/10^6$} & $T_{\rm fact}/10^9$ & \shortstack{Serial ref.\\$T_{\rm fact}/10^9$} & Change & \shortstack{Surface code\\(days)}\\
\midrule
1,435 & 8 & 1 & --- & 5.88 & --- & --- & 3.10\\
\midrule
1,403 & 8 & 1 & 466.32 & $4.2514\pm0.0031$ & 4.23 & $+0.6$\,\% & 3.31\\
1,450 & 8 & 2 & 327.90 & $2.9896\pm0.0011$ & 3.94 & $-24.0$\,\% & 2.66\\
1,685 & 8 & 8 & 188.90 & $1.7244\pm0.0007$ & 3.20 & $-46.2$\,\% & 2.07\\
1,796 & 8 & 12 & 170.51 & $1.5559\pm0.0009$ & 3.08 & $-49.5$\,\% & 1.66\\
2,050 & 8 & 16 & 148.12 & $1.3479\pm0.0009$ & 2.96 & $-54.4$\,\% & 1.18\\
\midrule
1,271 & 14 & 1 & 726.13 & $11.0210\pm0.0019$ & --- & --- & 7.26\\
1,861 & 5 & 8 & 228.49 & $1.3799\pm0.0005$ & 3.05 & $-54.7$\,\% & 1.14\\
2,358 & 4 & 19 & 195.43 & $0.9805\pm0.0004$ & 2.93 & $-66.6$\,\% & 1.17\\
2,432 & 4 & 22 & 187.60 & $0.9471\pm0.0005$ & 2.93 & $-67.7$\,\% & 1.13\\
2,998 & 3 & 37 & 206.22 & $0.8277\pm0.0004$ & 2.93 & $-71.8$\,\% & 1.08\\
\bottomrule
\end{tabular}

\caption{RSA-2048 resource estimates. The first row evaluates the settings of
Gidney's Table~5~\cite{gidney2025} in our benchmark, including the temporary
qubits and the address-determined simplifications of Sec.~\ref{sec:zero}.
The other rows give eight-replicate means with 95\,\% confidence half-widths.
Change is relative to the least costly serial model under the same qubit cap.
No serial setting fits at 1,271 qubits. Table~\ref{tab:all} lists all 42 circuits.}

\label{tab:summary}
\end{table}

At 1,403 qubits, a single-prime circuit costs 4.25 billion Toffolis, close
to the 4.23 billion available by retuning the serial model. At 1,450 qubits,
two primes can share a lookup. The measured mean falls to 2.99 billion
while the retuned serial value is 3.94 billion under that cap
(Table~\ref{tab:summary}). The gap that survives retuning is the quantity
we follow across the frontier.

\needspace{8\baselineskip}
\paragraph{Contributions.}
\begin{itemize}
\item \emph{Sharing across primes.} A single quantum table read returns the
  logarithm words of up to 37 primes, with their measurement phases merged
  into one correction (Sec.~\ref{sec:shared}, Fig.~\ref{fig:circuits}a).
\item \emph{Retained inputs and lazy residues.} Keeping the input of each
  multiplication replaces inverse modular arithmetic by subtraction and
  removes per-step reductions; the last multiplication feeds the CRT
  accumulator directly (Secs.~\ref{sec:ckpt}--\ref{sec:cross_stage},
  Fig.~\ref{fig:circuits}b). Appendix~\ref{app:proofs} proves that the core
  transformations preserve the reference output and its phases.
\item \emph{A budget-matched frontier.} Forty-two circuits between 1,271 and
  2,998 logical qubits, each scored in an open benchmark and replicated
  eight times; those that share lookups save 16--72\,\% against a serial model retuned under the
  same qubit cap (Fig.~\ref{fig:frontier}).
\item \emph{Physical consequences.} Within the surface-code, LDPC-cat, Pinnacle~1.0, and space-efficient neutral-atom models, the
  circuits shorten projected factoring time at a given footprint (whereas the balanced neutral-atom processor's best LP24-fitting design takes 29.3 years versus 25.4 years for its tuned reference),
  and expose a choice between the fewest gates, the least spacetime and the
  shortest time (Fig.~\ref{fig:arch}, Sec.~\ref{sec:meaning}).
\end{itemize}

\section{From an exponent to a masked residue}
\label{sec:background}

The quantum part of factoring must evaluate a modular product while
preserving interference between exponents with the same output. In the
Eker{\aa}--H{\aa}stad construction, a repetition parameter $s$ trades
input width for the number of runs~\cite{ekera2017}. For an $n$-bit
modulus, the two input registers $(e_1,e_2)$ together occupy
$m=\lceil n/2+n/(2s)\rceil+\lceil n/(2s)\rceil$ qubits and evaluate the two-base modular product $g^{e_1} h^{e_2}\bmod N$ with $h = g^{N-1}\bmod N$ as in Gidney's formulation~\cite{gidney2025}. At $n=2048$,
$s=8$ uses $1{,}280$ input qubits and nine nominal runs, whereas $s=3$
uses $1{,}708$ qubits and four runs. For resource counting, the benchmark substitutes a single-base proxy $g^e\bmod N$ over the same $m$-bit exponent split into $W_1$ windows of width $w_1$, which preserves the number of exponent windows and lookup-table dimensions (noted as a proxy in Sec.~\ref{sec:limitations}).

Divide the input into windows of width $w_1$. The value $e_j$ of window
$j$ selects a known integer factor $b_j(e_j)$ of the modular product.
For a single power $g^e$, for example,
$b_j(a)=g^{a2^{jw_1}}\bmod N$. The integer product of the selected
factors can be much larger than $N$. A residue number system evaluates
it modulo small auxiliary primes whose product is large enough for
reconstruction and close to a multiple of $N$~\cite{cfs2024,gidney2025}.

For each such prime $p$, fix a primitive root $g_p$. Loop~1 looks up
the discrete logarithms $d_{p,j}(e_j)$ of the selected factors modulo
$p$ and adds them. Loop~2 reduces their sum modulo $p-1$. Loop~3
converts the reduced logarithm back to a residue by windowed
exponentiation. Together these steps compute
\[
 S_p=\sum_j d_{p,j}(e_j),\qquad
 r_p=g_p^{\,S_p\bmod(p-1)}\bmod p
     =\Bigl(\prod_j b_j(e_j)\Bigr)\bmod p .
\]
Loop~4 adds the prime's weighted, truncated Chinese remainder theorem
(CRT) contribution to an $f$-bit accumulator modulo
$M_t=\lfloor N/2^{n-f}\rfloor$. The accumulator starts in a coherent
superposition of consecutive mask values. Each branch must acquire
the intended approximate offset and retain its relative phase after
the prime-specific registers are cleared.

For a small example, let $N=35$, $g=3$ and $e=6$. The two base-four
windows select $9$ and $11$, whose ordinary product is $99$. Modulo
$13$ the base-two logs are $8$ and $7$, so the residue is
$2^{15\bmod12}\bmod13=8$. Modulo $17$ the same two addresses return
base-three logs $2$ and $7$, giving $3^9\bmod17=14$. Both residues
describe the same integer $99$. What the primes can share is the
selection of the exponent window. Their logarithms and residues must
remain distinct. Appendix~\ref{app:construction} follows these values
through the two register lanes.

\section{Construction}
\label{sec:changes}

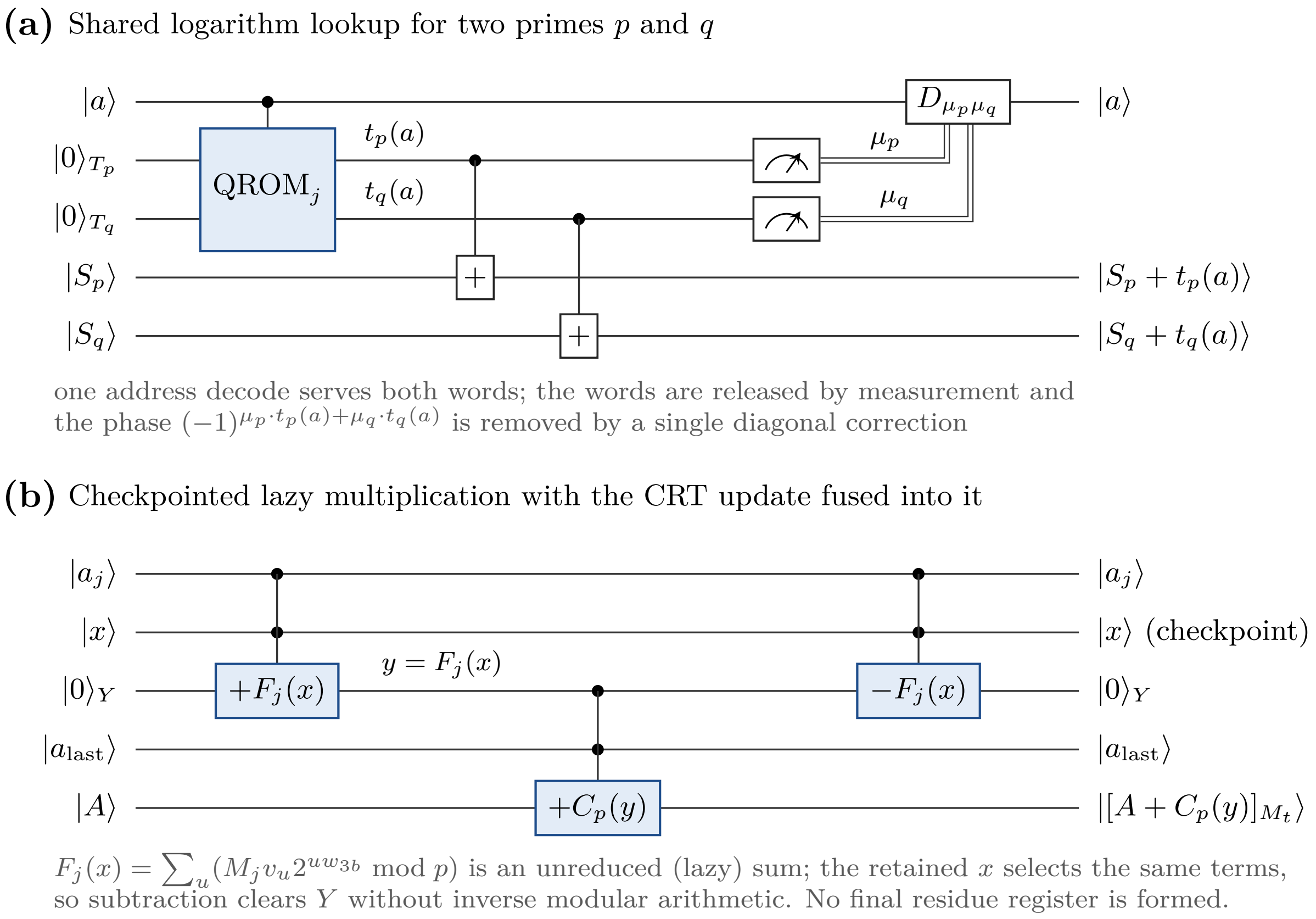
\begin{figure}[t]
\centering
\resizebox{\linewidth}{!}{\input{tikz_circuits}}
\caption{The two circuit fragments that carry the savings. (a) A single
QROM read at address $a=e_j$ returns a word for each of two primes; both
words are added to their own running logarithms and then released by
$X$-basis measurement. Double lines carry the outcomes to one diagonal
correction $D_{\mu_p\mu_q}$ on the address. (b) The retained checkpoint $x$
and selector $a_j$ determine $y=F_j(x)$. The last selector and $y$ address
the offset added to the accumulator modulo $M_t$, and subtracting $F_j(x)$
clears $Y$.}
\label{fig:circuits}
\label{fig:pipeline}
\end{figure}

Figure~\ref{fig:circuits} follows the temporary registers through a shared
lookup and through the last stages of a single prime's computation.

\subsection{One lookup for several primes}
\label{sec:shared}

Every prime uses the same window $e_j$ to address a different classical
table. Concatenating the corresponding entries for $k$ primes into one
row lets a single quantum read-only memory (QROM) operation return $k$
words, as in Fig.~\ref{fig:pipeline}(a). In the benchmark a full
$w_1$-bit read costs $2^{w_1}-w_1-1$ Toffolis irrespective of word
width. The group pays this address cost once and still pays for each
addition and its live register, or \emph{lane}. When a lane moves from
$p$ to $q$, each table supplies the difference
$d_{q,j}(e_j)-d_{p,j}(e_j)$. Summing the differences updates $S_p$
to $S_q$. A final move to an all-zero table clears the lane.

Measuring a temporary word in the $X$ basis releases it but introduces
the sign $(-1)^{\mu\cdot\mathbf t_j(a)}$ on address $a$, where $\mu$
is the measurement string. The signs for the same address multiply, so
the corresponding diagonal phase corrections can be combined and
deferred. When a packed row needs too many temporary qubits, the
circuit computes common Boolean products of the address bits and forms
one lane word at a time by XORs~\cite{li2026}.

Consider a group of eight primes and a seven-bit address, as in the
1,685-qubit circuit. For a full table the address selection costs
$2^7-7-1=120$ Toffolis. Eight separate reads would pay this cost eight
times. One packed read pays it once, although all eight additions and
the cleanup of their words still contribute to the circuit. The
saving on this one operation therefore cannot be applied to the full
factoring count.

\subsection{Retaining the input to a multiplication}
\label{sec:ckpt}
\label{sec:lazy}

Suppose a multiplier $M_j$ acts on an intermediate residue $x$. Erasing
$x$ early saves qubits, but reconstructing it during cleanup requires
inverse modular arithmetic. Keeping $x$ until the successor has served
its purpose trades register lifetime for a cheaper reversal, a reversible
time--space trade-off~\cite{bennett1989}. Write
$x=\sum_u v_u2^{uw_{3b}}$ in digits of width $w_{3b}$. Each digit
selects a table entry, and the successor register accumulates the
ordinary integer sum
\[
 F_j(x)=\sum_{u=0}^{W_3-1}
       \bigl(M_j v_u 2^{uw_{3b}}\bmod p\bigr)
       \equiv M_j x \pmod p .
\]
This \emph{lazy residue} can exceed $p$. Its calculation needs only
ordinary additions. Each of the $W_3=\lceil R/w_{3b}\rceil$ terms is
smaller than $p$, so a width $R$ with $W_3\le 2^{R-\ell}$ holds the
sum for any $\ell$-bit prime. During reversal, the retained $x$ selects
exactly the same terms, whose subtraction clears the register.
Appendix~\ref{app:construction} traces two consecutive multiplications
whose stored residues both exceed the prime.

\subsection{Compression and the final CRT update}
\label{sec:cross_stage}

Large groups also require space for the $k$ sums that persist between
windows. For $h=(p-1)/2$, a logarithm $qh+t$ contributes the sign
$(-1)^q$ and the smaller exponent $t$, since $g_p^h\equiv-1\pmod p$.
A fixed offset centres the sum of the $t$'s in a narrower signed
register. Its known contribution is restored in the residue lookup.
This compression, also studied by Li et al.~\cite{li2026}, can fail
in the tails of the sum distribution. The resource estimates deduct
an allowance extrapolated from sampled prime tables. A bound over an
entire prime set requires control of every prime
(Proposition~\ref{prop:mgf}).

The last multiplication ordinarily materialises a residue only for
loop~4 to split it into digits again. Instead, use a digit $v_u$ of
the live predecessor $y$ and the last multiplier selector $M$ as a
joint address. The table returns the truncated CRT offset for
$Mv_u2^{b_u}\bmod p$, where $b_u$ is the digit's starting bit.
Adding these offsets modulo $M_t$ gives the update $C_p(y)$ in
Fig.~\ref{fig:pipeline}(b). No final residue register is required.
Rounding each digit separately can change the last bits of the result.
We choose the partition to stay within the prime's existing number of
rounded additions and check the accumulated error against the benchmark
tolerance; because the branch verifier allows twice the per-addition deviation budget assumed in the keep-rate formula, passing sampled branch checks does not by itself certify the tighter keep-rate bound (Appendix~\ref{app:construction} states the worst-case modular deviation bound and the RNS capacity and nonzero-multiplier conditions). Appendix~\ref{app:construction} gives the register widths
and a worked digit partition. Appendix~\ref{app:proofs} establishes
output and phase preservation for the core, uncompressed transformations.

\section{Benchmark and verification}
\label{sec:bench}

We evaluate the circuits with Gidney's arithmetic simulator~\cite{gidney2025}.
For each primary score we run every auxiliary prime at 32, 64 and 128 bits
and at two settings of the 330-bit RSA-100 instance. On each sampled
basis branch the verifier checks the exponent, the unmasked output, the
release of every work register and the cancellation of relative phases.
The circuit must pay for any quantum-addressed table read. The peak
number of simultaneously live qubits defines $Q$.

At RSA-2048 we sample complete groups of primes to estimate their
operation counts, and include the partly filled last group at its actual
size. For the 1,685-qubit circuit, for example, the 18,931 primes fill
2,366 groups of eight and a final group of three. Appendix~\ref{app:bench}
gives the estimator and a worked calculation. Under the benchmark scoring convention, from the estimated count
per execution, $T_{\rm shot}$, we compute
\begin{equation}
 T_{\rm fact}=\frac{(s+1)T_{\rm shot}}{p_{\rm keep}-\eta_{\rm audit}} .
 \label{eq:fact}
\end{equation}
Here $s+1$ is the nominal run count from Eker{\aa}'s short-discrete-logarithm post-processing~\cite{ekera2020}, $p_{\rm keep}$ is the kept fraction
from masking and truncation in Gidney's bad-shot model~\cite{gidney2025}, and $\eta_{\rm audit}$ is the additional
allowance for overflow when logarithms are compressed. Equation~\eqref{eq:fact} is a benchmark scoring convention that treats $\eta_{\rm audit}$ as a reduction in the per-shot keep rate; operationally, a trace-distance bound alone (Proposition~\ref{prop:mgf}) does not justify this retry model for silent compression errors, and for the 30 sampled-allowance circuits $\eta_{\rm audit}$ remains conditional on the unexamined primes.
We did not re-evaluate Eker{\aa}'s classical post-processing on simulated outputs for the exact $s=3,4$ settings; the sensitivity is material—requiring one additional run at 2,998 qubits ($s+1=4$ to $5$) raises $T_{\rm fact}$ from 0.8277 to 1.035 billion Toffolis.
Twelve gap-fill circuits use all-prime exact integer MGF certificates, while the remaining 30 extrapolate from 27 selected primes per parameter set (Appendix~\ref{app:proofs}, Table~\ref{tab:all}).
Following the logical benchmark,
we omit a common 0.99 post-processing success factor on both sides of
the comparison.

\section{Results at RSA-2048}
\label{sec:results}

\begin{figure}[!t]
\centering
\includegraphics[width=\linewidth]{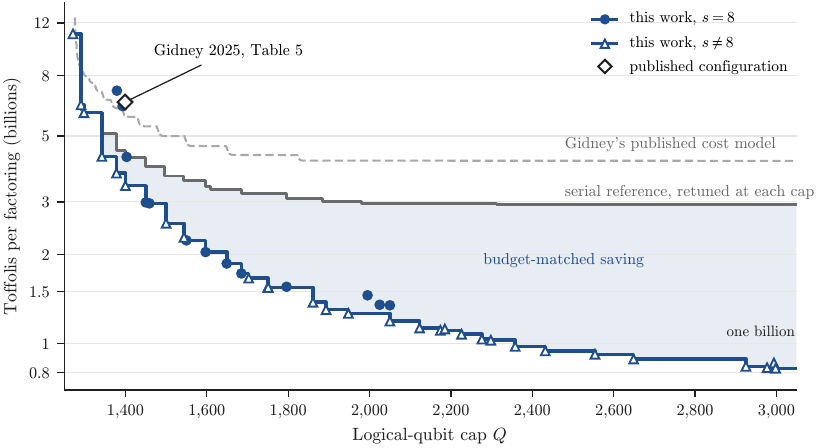}
\caption{Expected Toffolis per RSA-2048 factoring at a logical-qubit cap $Q$.
Dashed: least cost in Gidney's unmodified model at each cap; the diamond
is his Table~5 setting~\cite{gidney2025}. Grey: the serial reference, the
least costly single-prime ($k=1$) setting that fits the cap after its
arithmetic and parameters are retuned. Blue: the 42 circuits of this work
(eight-replicate means; circles $s=8$, triangles $s\neq8$) and the least cost
available under each cap. The shaded band is the saving that remains after
retuning. Table~\ref{tab:all} gives the 95\,\% intervals.}
\label{fig:frontier}
\end{figure}

\subsection{What additional logical qubits buy}
\label{sec:frontier}

Most of the 42 circuits in Fig.~\ref{fig:frontier} improve the best measured
cost as the cap increases, tracing a monotonic staircase from 1,271 to 2,998
logical qubits. A horizontal segment means that the same
circuit remains the least costly one available until a wider circuit
improves on it. Points above this staircase can still be the best choice
at a specified $s$ or on a particular machine. The two serial comparisons
use different arithmetic. Gidney's unmodified model gives a historical
reference with its own accounting. At its Table~5 parameters the model
gives 1,399 qubits and 6.50 billion Toffolis. Our benchmark counts
36 additional temporary qubits at those settings and gives 5.88 billion
after the address-determined simplifications (Table~\ref{tab:summary}).
The grey staircase shows $T_1(Q)$ from the primary serial model (\texttt{r3\_all\_generic\_with\_zero\_skip\_anf}, which supplies Tables~\ref{tab:summary} and~\ref{tab:all} and Fig.~\ref{fig:frontier}), and
Table~\ref{tab:all} lists the budget-matched saving $\Delta$ of each circuit at its own
peak. These 24.0\,\%--71.8\,\% figures compare circuits with jointly tuned $(s,k,\ell,w_1,f)$ against that primary serial model (3.9362\,B at 1,449 qubits, 2.9578\,B at 1,981 qubits, and 2.9321\,B at 2,312 qubits; by contrast, Table~\ref{tab:ablations}'s 30.7\,\% reduction for $k=3$ versus $k=1$ at fixed $\ell=21, w_1=6$ isolates lookup sharing alone). Under the expanded analytic sensitivity model (\texttt{r4b\_strictest\_k1\_bench}, which additionally credits $D=\ell+5$ compression and stricter generic arithmetic at $w_3=5, \ell\in\{20,21,22\}, f\in\{37,38\}$), winning serial estimates are 3.8336\,B at 1,448 qubits ($s=9,f=37$), 2.7128\,B at 1,831 qubits ($s=4,f=37$), 2.6496\,B at 2,004 qubits ($s=3,f=38$), and 2.6260\,B at 2,150 qubits ($s=3,f=38$), which still leaves substantial budget-matched savings of 22.0\,\%, 49.1\,\%, and 68.5\,\% at 1,450, 2,050, and 2,998 qubits; outside the 0.6\,\% check at 1,403 qubits, model error at other serial caps is unquantified. The serial search finds no feasible setting at the 1,271-qubit
cap. The budget-matched saving grows with the number of primes that share a lookup, from
16--25\,\% at $k=2$ to about two thirds for $k\ge19$. Single-prime circuits
at $s=8$ match the serial model within 1\,\% at 1,403 qubits and pay more
Toffolis below it in exchange for fewer qubits.

\paragraph{Keeping the repetition parameter fixed.}
When $s=8$, the exponent register takes 1,280 qubits across every circuit.
Raising the cap from 1,403 to 1,450 frees 47 logical qubits—just enough to
share each lookup across two primes instead of one—which drops the estimated
gate count by roughly 30\,\%, from 4.2514 to 2.9896 billion Toffolis. Under
the same 1,450-qubit cap, the best serial reference sits at 3.94 billion
Toffolis (using 1,449 qubits and $s=7$), leaving the 24.0\,\% saving shown
in Table~\ref{tab:summary}. Note that fixing $s$ in our candidate family
still leaves the serial comparator free to pick its own optimal repetition
parameter.

Expanding the budget further brings rapid gains at first: at 1,685 qubits,
eight primes share a lookup (1.7244 billion Toffolis, 46.2\,\% below the
3.203-billion serial reference), and at 1,796 qubits, twelve primes share
(1.5559 billion, a 49.5\,\% saving). Beyond that point, however, the returns
sharply diminish. Moving from 2,025 to 2,050 qubits at $s=8$ trims the
estimate from 1.3535 to 1.3479 billion (about 0.4\,\%) because both
circuits already group sixteen primes against a serial baseline near
2.958 billion; at fixed $s$, additional workspace has largely exhausted what
this window and group structure can buy.

\paragraph{Changing the number of runs.}
Varying $s$ opens a second trade-off between exponent width and run count.
At the space-constrained end ($s=14$), the exponent shrinks to 1,172 qubits,
allowing a complete circuit to fit inside 1,271 qubits at 11.021 billion
Toffolis—a regime where the serial grid has no feasible setting at all. At
the opposite end, smaller $s$ demands more exponent qubits but fewer
repetitions. At 2,050 qubits, for instance, the $s=8$ circuit needs about
148 million Toffolis per run compared with 235 million at $s=4$, yet
multiplying by nine runs versus five reverses their ranking (1.3479 versus
1.1930 billion Toffolis per factoring).

With $s=4$ and 19 primes per group at 2,358 qubits, the eight-replicate benchmark mean crosses below one billion for the first time ($0.9805\pm0.0004$ billion Toffolis) under the benchmark phase-table price, nominal $s+1$ run count, and omitted $0.99$ post-processing factor.
Under the constructive half-split phase-table price, however, the primary sample rises to $0.9971$ billion, and restoring the $0.99$ post-processing factor used in Gidney's Table~5 raises that figure to $1.0072$ billion.
Raising the group size to $k=22$ at 2,432 qubits lowers the benchmark mean to $0.9471\pm0.0005$ billion, achieving the smallest product $Q\,T_{\rm fact}$ ($2.303\times10^{12}$) in our set
and providing a stronger sub-billion milestone: its saved constructive-phase replicate mean is $0.9617$ billion, or $0.9714$ billion after restoring the $0.99$ factor,
still conditional on the nominal run count. The overall minimum gate count is
$0.8277\pm0.0004$ billion at 2,998 qubits ($s=3$, $k=37$), a 71.8\,\%
reduction below the 2.9321-billion serial baseline at that cap. Of course,
neither the logical Toffoli count nor $Q\,T_{\rm fact}$ accounts for the
physical time of each operation, which we turn to next.

\subsection{Which savings persist at fixed parameters}
\label{sec:path}

Because points along the frontier vary several parameters simultaneously,
Table~\ref{tab:ablations} isolates individual changes at matched settings.
Keeping four checkpoints at $\ell=21$, $w_1=6$ and $k=1$ saves 11.1\,\% in
Toffolis at the cost of 88 extra live qubits, and deferring reductions saves
a further 3.9\,\% for ten more qubits. With the same checkpoints and
canonical residues, sharing each lookup across three primes saves 30.7\,\%
relative to a single-prime lookup. Note that these contrasts have different
denominators; Appendix~\ref{app:ablation} records the exact settings and
counts for each pair.

\paragraph{Sampling precision.}\label{sec:uncertainty}
Each plotted circuit cost is the mean of eight independent sampled
estimates, with peak qubit counts identical across all eight replicates.
The largest relative 95\,\% confidence half-width is 0.067\,\% among the
shared circuits and 0.486\,\% across all 42 circuits, and every primary
estimate passed the 1\,\% complete-instance check at 330 bits (eight of
336 replicate cross-checks exceeded 1\,\%, all at the two narrow serial
settings, with a maximum discrepancy of 1.99\,\%). These intervals quantify
sampling variability under fixed arithmetic and keep-rate assumptions.

\section{Hardware projections}
\label{sec:arch}

\begin{figure}[!t]
\centering
\includegraphics[width=\linewidth]{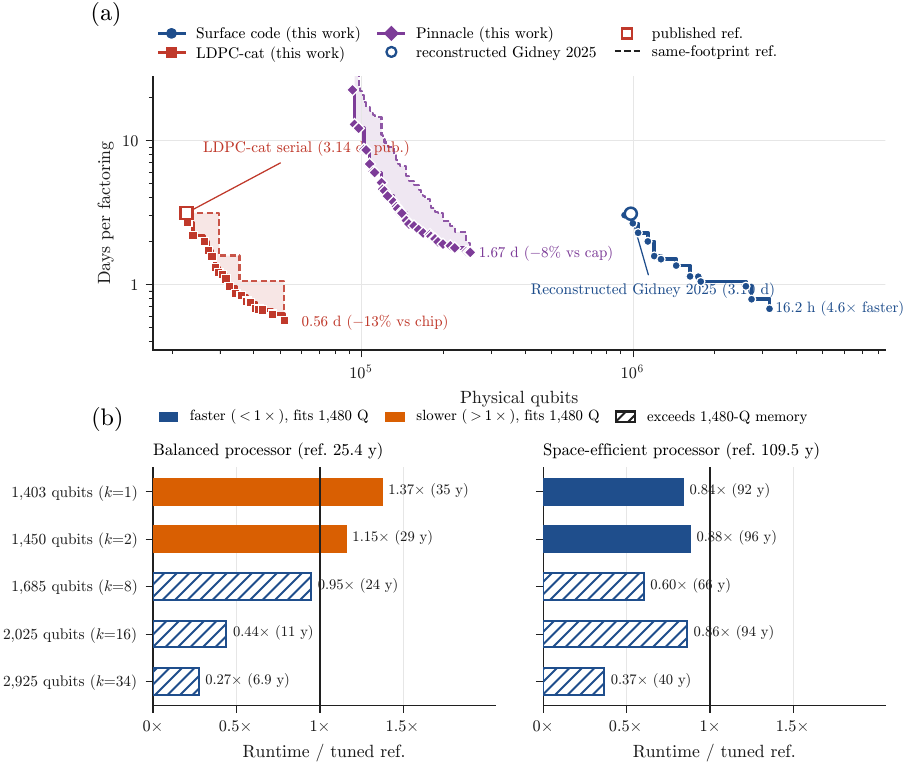}
\caption{Physical runtime projections across hardware models. (a) Physical
qubits versus days per factoring on the surface code, LDPC-cat, and Pinnacle~1.0.
Solid staircases and filled markers show the Pareto-optimal circuits of this
work; shaded bands highlight the speedup over the dashed same-footprint
references (Pinnacle~1.0 with retuned parallelism~\cite{pinnacle2026}, LDPC-cat
with its fastest listed chip~\cite{pappalardo2026}). Open symbols distinguish
the published LDPC-cat single-chip baseline~\cite{pappalardo2026} (3.14\,d) from this paper's surface-code reconstruction of Gidney's 2025 configuration (3.10\,d under our timing model, versus about 4.63\,d in~\cite{gidney2025}). (b)
Runtime relative to the tuned neutral-atom references of Cain et
al.~\cite{cain2026} (vertical $1\times$ line) at five representative logical
widths for the balanced processor variant (left, ref.\ 25.4\,y) and space-efficient
processor variant (right, ref.\ 109.5\,y); blue bars are faster ($<1\times$), orange
bars slower ($>1\times$), and hatched bars exceed the 1,480-qubit memory limit (extrapolated memory).
Appendix~\ref{app:transfer} lists all values.}
\label{fig:arch}
\end{figure}

A shared lookup selects an address once, yet the processor must still
deliver every word selected by that address. To estimate elapsed time, we
apply five published hardware models to the operation counts and word widths
of each circuit (Fig.~\ref{fig:arch}). Panel~(a) places each circuit at its
physical footprint alongside the fastest reference design of the same size,
whereas panel~(b) shows the relative runtime on Cain et al.'s neutral-atom
processors~\cite{cain2026} across representative qubit budgets against the
1,480-qubit memory limit.

\paragraph{Surface code.}
At a budget of one million physical qubits, our 1,450-logical-qubit circuit
takes 2.66 days on 997,192 physical qubits under our layout projection of Gidney's surface-code model~\cite{gidney2025} (leaving a 2,808-qubit margin under the cap, with routing allowance held fixed; Appendix~\ref{app:transfer}),
compared with 3.10 days for this paper's reconstruction of the fastest evaluated
serial window—about 14\,\% faster within our model (whereas Gidney's published timing calculation gives about 4.63 days). Allowing more active patches brings the shortest runtime
among the 42 circuits down to 16.2 hours (0.677 days) at 2,925 logical qubits (3.18
million physical qubits), whereas the gate-minimizing 2,998-qubit circuit
actually takes longer (25.9 hours, 1.079 days): in the surface-code timing model, \texttt{init\_lookup} scales as $2N/64$\,ms with address size $N$ (with no explicit output-width term), so although 2,998 qubits reduce addition time (\texttt{\_\_iadd\_\_}: 512 versus 7,337\,s/shot), wider lookup addresses drive \texttt{init\_lookup} time up from 5,229 to 20,188\,s/shot. Minimising
spacetime volume (physical qubits times days) selects a third operating
point at 1,861 logical qubits ($1.84$ million qubit-days, 39\,\% below
our reconstructed 2025 baseline; Fig.~\ref{fig:landscape}).

\paragraph{Neutral atoms.}
In the two processor variants of Cain et al.~\cite{cain2026}, a compact processor
reads from a larger memory: the balanced processor splits any lookup whose
output exceeds its working space and repeats part of the address selection,
whereas the space-efficient processor streams the output bits sequentially (separate from Cain et al.'s parallel 97-day architecture).
Eleven of our circuits fit inside the stated 1,480-qubit LP24 memory; oversized-memory points above 1,480 qubits in Table~\ref{tab:hw} are extrapolations. Among the circuits that fit,
the balanced processor reaches 29.3 years at 1,450 logical qubits (versus
25.4 years for its tuned reference), while the space-efficient processor
reaches 91.7 years at 1,403 qubits (versus 109.5 years for its reference). Thus, under the same memory limit, the two processor variants reach their shortest estimated times at different logical widths.

\paragraph{LDPC-cat and Pinnacle~1.0.}
Enlarging the work region also changes which serial or parallel reference
fits on the chip. On LDPC-cat~\cite{pappalardo2026}, the 2,925-qubit
circuit finishes in 0.561 days on 51,703 physical qubits when the serial
movement and measurement overhead is retained, compared with 3.14 days on
about 22,600 physical qubits for the published serial chip and 0.642 days
for the fastest published single-chip configuration within that larger
footprint (a 12.5\,\% same-budget saving). On Pinnacle~1.0~\cite{pinnacle2026} (arXiv:2602.11457v1; the later Pinnacle~1.2 release~\cite{pinnacle12} is not modelled here),
meanwhile, the 1,450-qubit circuit takes 6.05 days against a 14.99-day
reference at 109,860 physical qubits, and at 2,998 logical qubits the times
narrow to 1.67 versus 1.82 days at 250,800 physical qubits—because as the
physical cap grows, the candidate speeds up while the reference gains access
to greater parallelism. Appendix~\ref{app:transfer} gives the model
assumptions and per-circuit times.

\section{Discussion}
\label{sec:physical}

\subsection{Where the gate frontier flattens}

Increasing the group size initially removes repeated address selections from
the stage that dominates the serial gate count, but it cannot eliminate the
work of adding a distinct word to each prime's logarithm, converting each
logarithm back to a residue, or accumulating its CRT contribution. Once
those per-prime costs dominate, each additional lane must compete against
other uses for a qubit—such as extending the direct part of a lookup,
retaining an intermediate word, or supplying scratch space to shorten
cleanup—which explains why no single gate saving per extra lane holds across
Fig.~\ref{fig:frontier}.

At $s=3$, for example, the 2,925- and 2,998-qubit circuits differ by only
0.0127 billion Toffolis per expected factoring (about 1.5\,\% of the former),
because they already share across 34 and 37 primes, so adding three more
primes only reduces each prime's share of an address selection from $1/34$
to $1/37$. By contrast, their surface-code runtimes differ substantially,
showing that optimising for latency can matter far more than amortising the
address cost a little further.

Changing $s$ introduces two opposing effects: smaller values reduce the
number of runs the classical post-processing needs, yet each input is wider
and the prime system must cover a larger product (at 2,050 logical qubits,
for instance, the $s=4$ circuit requires 22,814 primes versus 16,532 at
$s=8$). Selecting a circuit for a finite machine therefore means balancing
run count, per-run cost, and the scratch space left over after the exponent
is allocated; the observed minimum at $s=3$ is specific to the circuits and
conventions tested here.

\subsection{What the estimates mean for RSA-2048}
\label{sec:meaning}

Table~\ref{tab:published} places the circuits next to published estimates
for the same task. In logical terms, sharing moves the cost of a residue-number-system factoring circuit
below one billion Toffolis at 2,358 logical qubits under the benchmark conventions ($0.9805$ billion, with constructive phase-table pricing plus the $0.99$ post-processing factor giving $1.0072$ billion,
while 2,432 logical qubits remain below one billion at $0.9714$ billion under both adjustments and the nominal run count), about 1.7 times the 1,399 qubits of Gidney's published
configuration, while every circuit from 1,500 qubits upward needs less than half
of that configuration's Toffoli count. In physical terms the surface-code
projection gives a choice that did not exist at a single operating point. A
machine with one million physical qubits factors in 2.66 days instead of
3.10. A machine three times larger, with 3.18 million physical qubits,
factors in 16.2 hours. For comparison, the 2019 construction of Gidney and
Eker{\aa} needed about 20 million physical qubits for an eight-hour
factoring~\cite{gidney2021}. On LDPC-cat hardware the fastest circuit takes
13.5 hours on 51,703 physical qubits, and on Pinnacle every circuit with a
same-footprint reference is faster than that reference
(Fig.~\ref{fig:arch}a).

These figures remain resource estimates rather than hardware demonstrations:
they depend on the benchmark's cost model, the sampled overflow allowance,
nominal repetition counts and published hardware models
(Sec.~\ref{sec:limitations}). While they do not alter the asymptotic
complexity of Shor's algorithm, they demonstrate that the constant factors
in the leading RNS construction were far from exhausted, and that its gate
count depends sensitively on how many qubits the arithmetic can borrow from
the exponent register.

\begin{table}[t]
\centering\small
\setlength{\tabcolsep}{4pt}
\begin{tabular}{@{}lrrrrl@{}}
\toprule
Construction & \shortstack{Logical\\qubits} & \shortstack{Toffolis\\($10^9$)} & \shortstack{Physical\\qubits} & Time & Accounting\\
\midrule
Gidney and Eker{\aa} 2019~\cite{gidney2021} & 6,189 & 2.62$^a$ & $2\times10^7$ & 8 h & published\\
Gidney 2025, Table~5~\cite{gidney2025} & 1,399 & 6.50 & $<10^6$ & $<1$ week & published model\\
\quad same settings in our benchmark & 1,435 & 5.88 & 0.98 M & 3.10 d & this benchmark\\
Li et al.\ 2026~\cite{li2026}, first $<10^9$ & 2,485 & 0.98 & --- & --- & reported, own\\
\midrule
This work, $s=8$ & 1,450 & 2.99 & 0.997 M & 2.66 d & this benchmark\\
This work, $s=8$ & 2,050 & 1.35 & 1.81 M & 1.18 d & this benchmark\\
This work, $s=4$, $<10^9$ (bench.$^b$) & 2,358 & 0.98 & 2.34 M & 1.17 d & this benchmark\\
This work, $s=4$, min $Q\,T_{\rm fact}$$^b$ & 2,432 & 0.95 & 2.44 M & 1.13 d & this benchmark\\
This work, $s=3$, fastest & 2,925 & 0.84 & 3.18 M & 16.2 h & this benchmark\\
This work, $s=3$, fewest Toffolis & 2,998 & 0.83 & 3.28 M & 25.9 h & this benchmark\\
\bottomrule
\end{tabular}
\caption{RSA-2048 estimates in context. Physical qubits and times for this
work use Gidney's surface-code model. $^a$Per modular exponentiation, from
the closed forms $3n+0.002n\lg n$ qubits and $0.3n^3+0.0005n^3\lg n$
Toffolis at $n=2048$. $^b$At 2,358 qubits, $0.9805\times 10^9$ is the benchmark replicate mean (omitting the $0.99$ post-processing factor); under constructive phase-table pricing ($0.9971\times 10^9$ on the primary sample) and restoring $0.99$, the figure becomes $1.0072\times 10^9$. At 2,432 qubits ($0.9471\times 10^9$ benchmark mean), the constructive-phase replicate mean is $0.9617\times 10^9$, or $0.9714\times 10^9$ after restoring $0.99$, still conditional on the nominal $s+1$ run count. Counts under different conventions (output
approximation, adder pricing, repetitions) are not directly comparable; the
budget-matched savings in this paper are measured only against the serial
model of Sec.~\ref{sec:frontier} within one benchmark.}
\label{tab:published}
\end{table}

\subsection{Relation to other constructions}
\label{sec:related}

Li et al.~\cite{li2026} cite an earlier preprint version of this work and already incorporate multi-prime lookup sharing, rolling lookup synthesis, and logarithm compression. Beyond their first sub-billion crossings at 2,433 logical
qubits (matched adder pricing) and 2,485 qubits (strict gate
counting), their Appendices~E.2--E.3 report 0.8986 billion strict-count Toffolis at 2,953 logical qubits.
Several techniques in our
cross-stage arithmetic draw directly on that work, including rolling
and bilinear algebraic-normal-form lookups, offset lane compression,
trailing-zero-sliced division and split phase trees. Section~\ref{sec:cross_stage}
locates each of these operations in the computation. Our fusion keeps the
truncated modulus $M_t$ used by the benchmark and carries the selected
multiplication digits directly into its CRT updates.
Proposition~\ref{prop:dyadic} explains why an accumulator change also
requires specifying how the mask is represented.

For numerical interpretation it matters that Li et al. use their own
gate accounting, approximation convention ($f$, dyadic accumulator), run-count assumptions, and compiled elementary gate sequences checked across all constituent primes, whereas our clearest contributions are (i)~the fixed-budget comparison against a retuned $k=1$ serial model, (ii)~the retained-input / lazy-residue construction under the benchmark's $M_t$ contract, and (iii)~the cross-architecture hardware trade-offs (Table~\ref{tab:li_compare}).
Neither set of
reported percentages supplies a head-to-head timing or gate comparison
between the two implementations.

\begin{table}[t]
\centering\small
\setlength{\tabcolsep}{4pt}
\begin{tabular}{@{}p{0.24\linewidth}p{0.35\linewidth}p{0.35\linewidth}@{}}
\toprule
Aspect & Li et al.\ (2026)~\cite{li2026} & This work\\
\midrule
Headline $<10^9$ / min $T_{\rm fact}$ & 2,485\,Q ($0.98$B); 2,953\,Q ($0.8986$B, strict) & 2,358\,Q ($0.9805$B bench.) / 2,432\,Q ($0.9714$B adj.); 2,998\,Q ($0.8277$B)\\
Accumulator \& mask & Dyadic CRT accumulator ($2^{32}/2^{33}$) & Benchmark $M_t=\lfloor N/2^{n-f}\rfloor$ ($f\in\{33,35,37\}$)\\
Gate pricing & Strict gate compilation + matched adder & Open benchmark + constructive $C_\phi(a)$ check\\
Verification scope & All-prime compiled gate checks & All-prime MGF on 12 gap-fill rows; 27-prime MGF + stratified sampling on 30\\
Primary contribution & LLM-guided compiler; strict gate counts & Budget-matched vs retuned $k=1$; lazy residues; hardware trade-offs\\
\bottomrule
\end{tabular}
\caption{Comparison of scope and conventions with Li et al.~\cite{li2026}.}
\label{tab:li_compare}
\end{table}

Other logical approaches change the factoring algorithm or the
representation of modular multiplication. Luongo et al.~\cite{luongo2025}
optimise windowed modular arithmetic. Regev's algorithm and its
space-efficient and streaming variants~\cite{regev2023,rv2024,vankirk2026}
have different quantum and classical workloads. On the hardware side,
Pinnacle, LDPC-cat and the neutral-atom proposals~\cite{pinnacle2026,pappalardo2026,cain2026,xue2026}
show how memory and processing choices alter the physical resources
needed for arithmetic. The comparisons in Sec.~\ref{sec:arch} use the
specified published models and do not assume that a new arithmetic
circuit can inherit a published routing schedule.

\section{Scope of the estimates}
\label{sec:limitations}

The core shared-lookup, checkpoint and lazy-residue transformations preserve
the coherent computation under the width conditions of
Proposition~\ref{prop:correct}, whereas CRT fusion changes where rounding
occurs and is verified against the allowed truncated output. For compressed logarithms, the bound in Proposition~\ref{prop:mgf} requires a
valid tail estimate for every prime. For eleven intermediate $(s, \ell, w_1)$
parameter sets we evaluate the float-free integer MGF bound across all
213,729 primes directly across 11 parameter sets using exact integer MGF certificates (where the RNS audit's \texttt{zero\_hits} field verifies that no window multiplier vanishes modulo an auxiliary prime, while the compression certificate reports a positive tail upper bound—for example, $\eta \le 3.18214\times 10^{-5}$ for $s=4,\ell=20,w_1=8$), whereas for the remaining
sets our reported allowance takes the worst of 27 evaluated primes per
parameter set; checking the remaining tables or proving a uniform bound
would remove that extrapolation. Two further
assumptions enter at the algorithmic level: the resource study models a
single-base arithmetic instance, and the $s=3$ and $s=4$ counts rely on the
nominal $s+1$ repetitions assumed by Eker{\aa}'s classical short-discrete-logarithm
post-processing~\cite{ekera2020}, which was not re-evaluated on simulated outputs for these settings (requiring one extra run at 2,998 qubits raises $T_{\rm fact}$ from 0.8277 to 1.035 billion).

The percentage savings in Table~\ref{tab:summary} also depend on the serial
search grid and its logical-qubit convention, which omits several scratch
qubits counted in our benchmark peak and therefore gives the serial baseline
extra room. Near the one-billion-gate threshold, phase-table synthesis
matters as well: pricing a constructive half-split phase lookup on the same
primary sample raises the 2,358-qubit estimate slightly from 0.9802 to
0.9971 billion Toffolis (Appendix~\ref{app:validation}). Finally, the
hardware projections rely on external assumptions about memory capacity,
qubit movement and routing, which are listed alongside the per-circuit times
in Appendix~\ref{app:transfer}.

\section{Conclusion}
\label{sec:conclusion}

The dominant costs in a residue-number-system factoring circuit come from
two repeated operations: decoding the same exponent address separately for
each prime, and rebuilding intermediate residues after discarding them.
Sharing the table lookup across primes and retaining multiplication inputs
attack both costs directly, while lazy residues and a fused final CRT
update trim the remaining per-prime arithmetic.

Under our keep-rate and compression assumptions, the $s=8$ circuits cut
expected Toffolis by 24.0\,\% at 1,450 logical qubits and by 54.4\,\% at
2,050 relative to the retuned serial baseline; allowing $s=3$ reaches
0.8277 billion Toffolis at 2,998 qubits. When mapped to physical hardware,
however, the best logical circuit is not always the best choice: on
Gidney's surface-code layout, 1,450 logical qubits fit below one million
physical qubits, 1,861 minimize spacetime volume under this timing model (1.8425 million qubit-days, about 0.8\,\% below the $s=8$ minimum at 2,025 qubits), and 2,925 minimize
runtime. A logical frontier bounds the arithmetic cost, but choosing an
operating point on real hardware still depends on the machine's memory
capacity and the relative price of its operations.

\paragraph{Data and code availability.}
\begingroup\sloppy
The benchmark, circuits, scores, replicate logs, tail certificates, primitive
tests and hardware scripts are archived at Zenodo,
\href{https://zenodo.org/records/23991044}{doi:10.5281/zenodo}.
Gidney's code is redistributed under CC~BY~4.0.
\par\endgroup

\paragraph{Acknowledgements.}
This research is supported by the National Research Foundation, Singapore,
through the National Quantum Office, hosted in A*STAR, under its Quantum
Engineering Programme~3.0 Funding Initiative (W24Q3D0002). Any opinions,
findings, conclusions, or recommendations expressed in this material are those
of the authors and do not reflect the views of the National Research
Foundation, Singapore. The benchmark, circuits and analysis were developed with
an AI-assisted multi-agent research workflow; all numerical results come from
the unmodified benchmark, exact certificates and the stated hardware models.
We thank Craig Gidney for publishing his code and Joshua~M.\ Courtney (Zapata
Quantum Inc.) for pointing out the redundant comparison that led to the
simplifications of Sec.~\ref{sec:zero}.

\bibliographystyle{unsrt}
\begingroup\small\raggedright
\bibliography{references}
\endgroup

\clearpage
\input{appendix}

\end{document}

%% file: numbers.tex
\newcommand{\baseQ}{1{,}435}
\newcommand{\baseTshot}{6.46\times10^{8}}

\newcommand{\trimOnlyQ}{1{,}432}
\newcommand{\trimOnlyTshot}{6.24\times10^{8}}

\newcommand{\trimOnlydT}{$-$3.3}

\newcommand{\trimOnlydQT}{$-$3.5}

\newcommand{\trimCkptQ}{1{,}520}
\newcommand{\trimCkptTshot}{5.55\times10^{8}}

\newcommand{\trimCkptdT}{$-$14.1}

\newcommand{\trimCkptdQT}{$-$9.0}

\newcommand{\lazyOnlyQ}{1{,}530}
\newcommand{\lazyOnlyTshot}{5.33\times10^{8}}

\newcommand{\lazyOnlydT}{$-$17.5}

\newcommand{\lazyOnlydQT}{$-$12.0}

%% file: frontier_numbers.tex
\newcommand{\frontierGrid}{534{,}240}

%% file: tikz_overview.tex
% Figure 1: the idea of the paper. Requires tikz with libraries positioning, calc, decorations.pathreplacing.
\begin{tikzpicture}[
  font=\small, >=stealth,
  dec/.style={draw=decc, fill=decf, rounded corners=1.5pt, minimum height=5.2mm, minimum width=15mm, inner sep=1pt, font=\footnotesize},
  word/.style={draw=wordc, fill=wordf, minimum height=5.2mm, minimum width=6.5mm, inner sep=0pt, font=\scriptsize},
  addb/.style={draw=addc, fill=addf, circle, inner sep=0.6pt, font=\scriptsize},
  lab/.style={font=\footnotesize, text=black!75},
  stage/.style={draw=black!45, fill=black!4, rounded corners=2pt, minimum height=11mm, text width=26mm, align=center, font=\footnotesize},
  mech/.style={draw=wordc, fill=wordf!60, rounded corners=2pt, text width=26mm, minimum height=10mm, align=center, font=\scriptsize, inner sep=2pt},
]
% ---------------------------------------------------------------- (a) one prime per lookup
\node[anchor=west, font=\normalsize\bfseries] at (-0.1,3.05) {(a)};
\node[anchor=west, lab] at (0.55,3.05) {Serial: one prime per lookup};
\foreach \p/\x in {1/0, 2/2.95, 3/5.9}{
  \node[dec] (d\p) at (\x+0.65,2.35) {decode $e_j$};
  \node[word, right=1.5mm of d\p] (w\p) {$t_{\p}$};
  \node[addb, right=1.5mm of w\p] (a\p) {$+$};
  \node[font=\scriptsize, below=0.5mm of a\p, text=black!70] {$S_{p_\p}$};
}
\draw[decorate, decoration={brace, mirror, amplitude=3pt}] (0.1,1.72) -- (8.55,1.72)
  node[midway, below=3pt, lab] {address decoding paid $k$ times: \ $k\,(2^{w_1}-w_1-1)$ Toffolis};
% ---------------------------------------------------------------- (b) k primes per lookup
\node[anchor=west, font=\normalsize\bfseries] at (9.15,3.05) {(b)};
\node[anchor=west, lab] at (10.15,3.05) {This work: $k$ primes per lookup};
\node[dec] (D) at (10.2,2.35) {decode $e_j$};
\node[word, right=1.5mm of D] (W1) {$t_1$};
\node[word, right=0mm of W1] (W2) {$t_2$};
\node[word, right=0mm of W2] (W3) {$t_3$};
\foreach \p/\dx in {1/0.0, 2/0.62, 3/1.24}{
  \node[addb] (A\p) at ($(W3.east)+(0.55+\dx,0)$) {$+$};
  \node[font=\scriptsize, below=0.5mm of A\p, text=black!70] {$S_{p_\p}$};
}
\draw[->, black!55, thin] (W3.east) -- (A1.west);
\draw[decorate, decoration={brace, mirror, amplitude=3pt}] (9.35,1.72) -- (14.75,1.72)
  node[midway, below=3pt, lab] {paid once: \ $2^{w_1}-w_1-1$ Toffolis};
% ---------------------------------------------------------------- (c) where the mechanisms act
\node[anchor=west, font=\normalsize\bfseries] at (-0.1,0.55) {(c)};
\node[anchor=west, lab] at (0.55,0.55) {Per-prime pipeline of the reference circuit and the changes made here};
\foreach \i/\t/\s in {1/{Loop 1\\discrete logs, 61\,\%}/l1, 2/{Loop 2\\reduce mod $p-1$}/l2,
                     3/{Loop 3\\residue $g_p^{d}$, 10\,\%}/l3, 4/{Loop 4\\CRT into $\mathbb{Z}_{M_t}$}/l4,
                     5/{Cleanup\\23\,\% in loop 3}/l5}{
  \node[stage] (\s) at ({(\i-1)*3.05+1.45},-0.35) {\t};
}
\foreach \a/\b in {l1/l2, l2/l3, l3/l4, l4/l5}{\draw[->, black!55] (\a) -- (\b);}
\node[mech, below=4mm of l1] (m1) {shared lookups\\compressed lanes\\monomials once per window};
\node[mech, below=4mm of l2] (m2) {half-modulus quotient\\signed long division};
\node[mech, below=4mm of l3] (m3) {checkpoints\\lazy residues};
\node[mech, below=4mm of l4] (m4) {fused with loop 3\\two-chunk update};
\node[mech, below=4mm of l5] (m5) {subtract kept terms\\merged phase fixes};
\foreach \i in {1,...,5}{\draw[wordc, thick] (l\i.south) -- (m\i.north);}
\node[font=\scriptsize, text=black!60, anchor=north west] at (-0.05,-2.75)
  {Percentages: share of reference Toffolis at the published settings. Section~\ref{sec:changes} describes each change.};
\end{tikzpicture}

%% file: tikz_circuits.tex
% Figure 2: circuits for a shared lookup and for a checkpointed multiplication fused with the CRT update.
\begin{tikzpicture}[
  font=\small, >=stealth, x=1cm, y=0.62cm,
  wire/.style={black!80, line width=0.55pt},
  gate/.style={draw=black!85, fill=white, line width=0.6pt, minimum height=4.6mm, inner sep=2pt, font=\footnotesize},
  big/.style={draw=wordc, fill=wordf, line width=0.7pt, font=\footnotesize},
  ctrl/.style={circle, fill=black, inner sep=0pt, minimum size=3.6pt},
  cls/.style={double distance=1.1pt, black!80, line width=0.4pt},
]
\newcommand{\meter}[2]{% node name, position
  \node[gate, minimum width=7mm] (#1) at #2 {};
  \draw[line width=0.5pt] ([xshift=-2.2mm,yshift=-1.0mm]#1.center) arc (160:20:2.35mm);
  \draw[->, line width=0.5pt] ([yshift=-1.3mm]#1.center) -- ++(55:2.6mm);}
% ---------------------------------------------------------------- (a) shared lookup
\node[anchor=west, font=\normalsize\bfseries] at (-1.55,1.3) {(a)};
\node[anchor=west, font=\footnotesize] at (-0.85,1.3) {Shared logarithm lookup for two primes $p$ and $q$};
\foreach \y/\l/\r in {0/{$\ket{a}$}/{$\ket{a}$}, -1/{$\ket{0}_{T_p}$}/{}, -2/{$\ket{0}_{T_q}$}/{},
                      -3/{$\ket{S_p}$}/{$\ket{S_p+t_p(a)}$}, -4/{$\ket{S_q}$}/{$\ket{S_q+t_q(a)}$}}{
  \node[anchor=east, font=\footnotesize] at (-0.05,\y) {\l};
  \node[anchor=west, font=\footnotesize] at (10.05,\y) {\r};
}
\draw[wire] (0,0) -- (10,0);
\draw[wire] (0,-1) -- (6.55,-1);  \draw[wire] (0,-2) -- (6.55,-2);
\draw[wire] (0,-3) -- (10,-3);    \draw[wire] (0,-4) -- (10,-4);
\draw[wire] (1.4,0) -- (1.4,-0.6);
\node[ctrl] at (1.4,0) {};
\node[big, minimum width=13mm, minimum height=13mm] (qrom) at (1.4,-1.5) {$\mathrm{QROM}_j$};
\node[font=\scriptsize, above] at (2.75,-1) {$t_p(a)$};
\node[font=\scriptsize, above] at (2.75,-2) {$t_q(a)$};
% additions: word register controls an adder onto S
\node[ctrl] at (3.6,-1) {}; \draw[wire] (3.6,-1) -- (3.6,-2.75);
\node[gate] at (3.6,-3) {$+$};
\node[ctrl] at (4.7,-2) {}; \draw[wire] (4.7,-2) -- (4.7,-3.75);
\node[gate] at (4.7,-4) {$+$};
% measure the words in the X basis
\meter{m1}{(6.9,-1)}
\meter{m2}{(6.9,-2)}
\draw[cls] (m1.east) -- (8.6,-1) -- (8.6,-0.3);
\draw[cls] (m2.east) -- (8.85,-2) -- (8.85,-0.3);
\node[gate, minimum width=11mm] at (8.72,0) {$D_{\mu_p\mu_q}$};
\node[font=\scriptsize] at (7.95,-0.68) {$\mu_p$};
\node[font=\scriptsize] at (8.05,-1.68) {$\mu_q$};
\node[font=\scriptsize, text=black!65, align=left, anchor=north west] at (-1.0,-4.55)
  {one address decode serves both words; the words are released by measurement and\\
   the phase $(-1)^{\mu_p\cdot t_p(a)+\mu_q\cdot t_q(a)}$ is removed by a single diagonal correction};
% ---------------------------------------------------------------- (b) checkpointed multiplication and fused CRT
\begin{scope}[yshift=-5.0cm]
\node[anchor=west, font=\normalsize\bfseries] at (-1.55,1.3) {(b)};
\node[anchor=west, font=\footnotesize] at (-0.85,1.3) {Checkpointed lazy multiplication with the CRT update fused into it};
\foreach \y/\l/\r in {0/{$\ket{a_j}$}/{$\ket{a_j}$}, -1/{$\ket{x}$}/{$\ket{x}$ (checkpoint)}, -2/{$\ket{0}_Y$}/{$\ket{0}_Y$},
                      -3/{$\ket{a_{\rm last}}$}/{$\ket{a_{\rm last}}$}, -4/{$\ket{A}$}/{$\ket{[A+C_p(y)]_{M_t}}$}}{
  \node[anchor=east, font=\footnotesize] at (-0.05,\y) {\l};
  \node[anchor=west, font=\footnotesize] at (10.05,\y) {\r};
  \draw[wire] (0,\y) -- (10,\y);
}
\node[ctrl] at (1.5,0) {}; \node[ctrl] at (1.5,-1) {}; \draw[wire] (1.5,0) -- (1.5,-1.72);
\node[big, minimum width=13mm] at (1.5,-2) {$+F_j(x)$};
\node[font=\scriptsize, above] at (3.25,-2) {$y=F_j(x)$};
\node[ctrl] at (4.9,-2) {}; \node[ctrl] at (4.9,-3) {}; \draw[wire] (4.9,-2) -- (4.9,-3.72);
\node[big, minimum width=13mm] at (4.9,-4) {$+C_p(y)$};
\node[ctrl] at (8.3,0) {}; \node[ctrl] at (8.3,-1) {}; \draw[wire] (8.3,0) -- (8.3,-1.72);
\node[big, minimum width=13mm] at (8.3,-2) {$-F_j(x)$};
\node[font=\scriptsize, text=black!65, align=left, anchor=north west] at (-1.0,-4.55)
  {$F_j(x)=\sum_u (M_j v_u 2^{u w_{3b}} \bmod p)$ is an unreduced (lazy) sum; the retained $x$ selects the same terms,\\
   so subtraction clears $Y$ without inverse modular arithmetic. No final residue register is formed.};
\end{scope}
\end{tikzpicture}

%% file: appendix.tex
% Appendices of main.tex (included with \input{appendix} after the bibliography).
\appendix
\section{Worked circuits and register accounting}
\label{app:construction}

\subsection{Tracing two exponent windows across two primes}

Take $N=35$, base $g=3$, and exponent $e=6$. In base four the two
digits are $e_0=2$ and $e_1=1$. The selected factors are $b_0=9$ and
$b_1=11$, since $3^4\equiv11\pmod{35}$. Thus the integer product is
$99$, and the desired modular product is $29$. We can check the first
three loops using the auxiliary primes $13$ and $17$. This example
illustrates the register arithmetic before CRT reconstruction.

Choose base $2$ for the discrete logarithms modulo $13$ and base $3$
modulo $17$. The table rows selected by the two window values are
$\mathbf t_0(2)=(8,2)$ and $\mathbf t_1(1)=(7,7)$. The first row follows
from $2^8\equiv9\pmod{13}$ and $3^2\equiv9\pmod{17}$. The second
follows from $2^7\equiv11\pmod{13}$ and $3^7\equiv11\pmod{17}$.
The register values after each packed addition are therefore
\[
 (0,0)\ \longrightarrow\ (8,2)\ \longrightarrow\ (15,9).
\]
Reducing the two entries modulo $12$ and $16$ gives $(3,9)$.
Exponentiating in the corresponding bases returns $(8,14)$, which is
$(99\bmod13,99\bmod17)$. The two prime lanes follow different
arithmetic from the moment the words are added. Only the lookup
address has to be decoded jointly.

A single lane can also move between primes. Suppose it currently holds
$15$, the sum for $p=13$, and will next serve $q=17$. The two difference
entries are $2-8=-6$ and $7-7=0$. Adding them leaves $9$, the
unreduced sum for $q$. Moving from $q$ to the zero row subtracts $2$
and $7$, and clears the lane. For a group of $k$ primes the same
telescoping takes place in each lane, with consecutive primes in that
lane separated by $k$ positions in the prime list.

\subsection{Table width, signs and division}
\label{sec:refine}

At $m=1{,}280$, widening an exponent window from six to eight bits
reduces the number of windows from $214$ to $160$. The price of a full
lookup rises from $2^6-6-1=57$ Toffolis to $2^8-8-1=247$. All $k$
lanes see the reduction in the number of additions. A single address
decoder serves the group at each window. The last table has only the
remaining address bits. For $21$-bit primes the full-period sums in
both examples fit in $29$ bits, from the bound
$S_p<W_1(2^\ell-1)$. Trimming a $32$-bit sum to that width saves three
qubits per lane and three high-order positions in the reduction.

A difference between two full-period table entries fits in $\ell+1$
signed bits. A biased unsigned encoding can alternate the sign of
successive updates so that the biases cancel in pairs, with a known
correction for an unpaired term. A different width saving comes from the
half-period $h=(p-1)/2$. Write each full-period table entry as $qh+t$,
where $q\in\{0,1\}$ and $0\leq t<h$. For $p=13$, base $2$ and
$d=9$, one obtains $q=1$, $t=3$, and $2^9\equiv-2^3\equiv5\pmod{13}$.
Across a full set of windows the circuit accumulates the $t$'s and
keeps the parity of the individual $q$'s. Reducing the sum of the $t$'s
modulo $h$ supplies a further quotient parity. The XOR of the two
parities determines the sign of the residue.

For a $20$-bit prime, the $19$-bit half-period remainder can be split
into an $11$-bit direct lookup and two four-bit multiplier selectors.
A signed-digit implementation of long division combines restoration
after a failed trial with the next trial at the halved divisor. Their
net value is one signed update, whose sign follows from the preceding
trial. Only the final remainder may need restoration. Trailing zero
bits of each shifted divisor can be omitted from that addition.

In a centred lane a fixed offset is subtracted from the half-period
sum. Its contribution is folded into the table for the initial
exponentiation. With $m=1{,}280$, $w_1=7$ and $\ell=20$, there are
$183$ windows. The worst-case sum fits in $27$ bits. A $25$-bit
centred lane releases $32$ persistent qubits in a group of $16$.
The signed divider accepts a finite interval about the chosen offset.
Proposition~\ref{prop:mgf} gives the interval, the tail bound for the
actual window tables and the conversion from lane failure to a per-run
allowance. Offset centring and shorter division slices are also used by
Li et al.~\cite{li2026}.

\paragraph{Producing a word when its lane is ready.}
For three address bits $(a,b,c)$, the possible nonlinear monomials are
$ab$, $ac$, $bc$, and $abc$. For instance, output bits with Boolean
forms $a\oplus bc$ and $ac\oplus abc$ can both be obtained by CNOTs
from the address and product registers. More generally, there are $2^w-w-1$
nonlinear monomials on a $w$-bit address. A lookup can populate the
whole register once. Each lane then forms the word specified by its
classical XOR coefficients, adds the word, and releases it by an
$X$-basis measurement. A sign bit can be accumulated directly into the
parity register. After the last lane, the product register is measured,
allowing its phase to be combined with the other phases for the same
address. The group still needs $k$ persistent sums, but only a single
temporary lane word. For short groups, the packed read can occupy less
space. Bilinear decompositions also allow products whose coefficients
vanish to be omitted~\cite{li2026}.

A large direct table can be split into banks selected by the high
address bits. The low bits select a base word and XOR differences
between banks. The high address bits determine which differences are
folded into the result. The cost model counts the additional temporary
words and their phase cleanup as well as the read~\cite{low2018}.
The same direct lookup can replace a multiplication when the combined
address remains short.

\subsection{Tracing a lazy multiplication and its erasure}

For $p=13$ use base-four digits and start from $x=7=3+4$. For a
multiplier of $3$, the two digit lookups return $9$ and $12$. Ordinary
addition stores $21$. The result represents $8\bmod13$. To multiply
again, this time by $2$, read the digits of the stored $21=1+4+16$.
Their contributions modulo $13$ are $2$, $8$ and $6$, giving the next
register value $16\equiv3\pmod{13}$. Reversal with the retained $21$
subtracts $6$, $8$ and $2$ from $16$, returning the target to zero.
Retaining $7$ lets the preceding step subtract $12$ and $9$ from $21$
in turn. This trace illustrates why each reverse lookup must see the
digits of the particular representative used in the forward circuit.

At RSA-2048 widths, a $24$-bit lazy word at $\ell=21$ can be split
into eight three-bit digits. Each contributes less than $p$, and
$8p<2^{24}$, so even the unreduced sum fits. The same bound applies to
the next multiplication. If a subsequent table requires a canonical
residue, the circuit includes a final reduction and its reversal.
A table defined on all the digits of the lazy word can consume the
stored representative directly.

\subsection{Direct writes and cleanup}
\label{sec:zero}

A zero target changes the first term of an addition into a table write.
For a modular subtraction from zero, the target and borrow are both
functions of the address, so both can be tabulated. The same argument
applies to the start of a canonical rebuild and to the final erasure
of its known term. A lazy accumulation has an analogous first write
and last deletion. With three-bit exponent windows, the residue after
three windows depends on nine bits and fits a single $512$-entry table.
This replaces a direct lookup of two windows and the first subsequent
multiplication. Since the resulting source is canonical, source digits
entirely above its $\ell$ value bits vanish. These direct writes and
lookup choices also enter the serial reference. At Gidney's published
settings they reduce the benchmark count by $10.2\,\%$.

\subsection{The number and width of CRT offsets}

Let $y=\sum_u v_u2^{b_u}$ be the live predecessor of the final
multiplication, let $M$ be the last selected multiplier and let
$P_*=\prod_{p\in P}p$. Write $c_p=(P_*/p)^{-1}\bmod p$. In the exact
CRT sum, a single digit contribution $t_u=M v_u2^{b_u}\bmod p$ has
coefficient $(c_p t_u)\bmod p$ multiplying $P_*/p$. If the contributions
are accumulated separately, the sum of their coefficients may exceed
the coefficient of the combined residue by a multiple of $p$. With
$d$ digits it exceeds it by at most $(d-1)p$, since each coefficient lies
in $[0,p)$. The resulting exact CRT contributions differ by a multiple
of $P_*$. The reference selects $P_*$ close to a multiple of $N$, and
the offset tables also truncate each digit contribution. Thus both the
number of digit contributions and the chosen modulus matter when
comparing the approximate outputs.

In the $\ell=19$ audit, a $22$-bit predecessor is partitioned into
widths $6+6+5+5$ with starting positions $0,6,12,17$. Each table uses
the multiplier selector and one of those digits as its address. The
four adds fit the existing four-offset allowance. The audit also checks
the circular distance of the accumulated value from the exact CRT
contribution at the sampled inputs.

A different constraint arises when the temporary offset word sets the
qubit peak. For $f=33$ a subtraction can read its low $14$ bits,
subtract them with carry propagation, and release that word before
reading and subtracting the high $19$ bits at bit position $14$.
These are two parts of the same integer offset. Slicing therefore
adds no new rounding step. Both lookups depend on the same digit and
selector, allowing their measurement phases to be combined before the
common correction. Modular restoration follows the release of the two
temporary words. Other circuits alternate the signs used in successive
digit tables to reuse an available comparison word during cleanup.
Compact ripple-carry adders~\cite{cuccaro2004} and sign-cleanup
schedules give different gate counts at these small qubit budgets. \paragraph{Deviation budget, RNS capacity, and nonzero multipliers.} For a fused CRT stage with at most $d\le 5$ digit additions per prime across $|P|$ primes, accumulator width $f$, and wraparound gap $g$, the keep-rate formula budgets a modular deviation of $d|P|(2\cdot 2^{-f}+2^{-g})$, whereas the benchmark's branch verifier (\path{rsabench/instances.py}) permits twice that budget, $2d|P|(2\cdot 2^{-f}+2^{-g})$. For example, at $Q=2,358$ ($f=37, g=37, |P|=19{,}881, d\le 5$), the keep-rate deviation budget is at most $2.17\times 10^{-6}$ versus a verifier bound of $4.34\times 10^{-6}$ (with a 28-bit mask giving mask proportion $2^{28-37}=1.953125\times 10^{-3}$). Passing sampled branch checks therefore confirms modular correctness within the $2\times$ tolerance without independently certifying the tighter keep-rate budget. All RNS caches satisfy the reconstruction-capacity condition $P_\ast \ge N^{W_1}$ for RSA-2048 (for example, $450{,}595$ bits for the certified $s=4,\ell=20,w_1=7$ configuration), and every RNS audit verifies \texttt{zero\_hits}=0 so that no addressed window multiplier vanishes modulo any auxiliary prime where discrete-logarithm tables are used.

\begin{table}[htbp]
\centering\small
\begin{tabular}{lp{0.67\linewidth}}
\toprule
Symbol & Meaning\\
\midrule
$s$, $m$ & Repetition parameter and combined input width. A factoring
requires $s+1$ nominal runs.\\
$P$, $\ell$, $k$ & Auxiliary prime set, prime bit length and number of
primes sharing an exponent-window lookup.\\
$w_1$, $W_1$ & Width and number of input windows.\\
$L$, $D$, $R$ & Deterministic logarithm-sum width, compressed lane width
and lazy-residue width.\\
$h$, $g_p$ & Half-period $(p-1)/2$ and primitive root modulo $p$.\\
$w_{3a}$, $w_{3b}$, $w_4$ & Widths of the multiplier selector, source
digit and CRT digit in the unfused calculation.\\
$f$, $M_t$ & Accumulator width and truncated modulus.\\
$p_{\rm keep}$, $\eta_{\rm audit}$ & Reference keep rate and extrapolated
compression allowance used in scoring.\\
$Q$, $T_{\rm shot}$, $T_{\rm fact}$ & Peak logical qubits, Toffolis per
execution, and expected Toffolis per factoring.\\
\bottomrule
\end{tabular}
\caption{Symbols used in the construction and resource estimates.
Register widths are in bits.}
\label{tab:notation}
\end{table}

\section{Output preservation, tail bounds and the CRT modulus}
\label{app:proofs}

\begin{proposition}[Output preservation for the core transformations]\label{prop:correct}
Fix the modulus and generator, $s$, the prime size
$\ell$ and the RNS primes with their tables, the windows $w_1$, $w_3$ and
$w_4$, the accumulator width $f$ and the mask of the reference configuration.
Assume $L\le32$ for the source
tables represented modulo $2^{32}$ (\ref{app:proofs}1), and that the chosen widths satisfy
$W_1\le2^{L-\ell}$ (trimmed register, \ref{app:proofs}2) and
$\lceil R/w_{3b}\rceil\le2^{R-\ell}$ (lazy register, \ref{app:proofs}3). Then, for every
exponent $e$ and every outcome of the intermediate X-basis measurements, the
circuit with shared loop-1 lookups (any $k$), the trimmed register,
checkpoints, lazy residues and the address-determined steps of
Sec.~\ref{sec:zero}
\begin{enumerate}
\item leaves the exponent register unchanged,
\item produces the same masked approximate output as the reference circuit for
  that configuration,
\item returns every work register to zero, and
\item leaves no input-dependent relative phase after the deferred phase
  corrections,
\end{enumerate}
so that, after the measurement transcript is discarded, it implements the same
isometry as the reference circuit on arbitrary coherent inputs (\ref{app:proofs}5).
\end{proposition}

\noindent\emph{Proof.}
We first track the register values for a basis-state exponent (\ref{app:proofs}1--\ref{app:proofs}4),
then track the relative phases required for a coherent superposition (\ref{app:proofs}5).
The final substitution replaces two table-based operations with a single
table encoding their composition (\ref{app:proofs}6). Lane compression is treated
separately in Proposition~\ref{prop:mgf}.

\paragraph{\ref{app:proofs}1. Shared-register invariant.}
The reference implementation stores loop-1 tables as differences between
consecutive primes, modulo $2^{32}$, with a final row that returns the running
sum to zero. Let $C_i(j,e)$ be the cumulative sum of rows $0,\dots,i$. Thus
$C_i(j,e)\equiv d_{i,j}(e)\pmod{2^{32}}$ and $C_{|P|}(j,e)\equiv0$. Register $x$
of a group follows the chain of primes $x,x+k,x+2k,\dots$. When it moves from
prime $a$ to prime $b$, the shared lookup adds
$C_b(j,e_j)-C_a(j,e_j) \bmod 2^{L}$ for every window $j$ (with $C_{-1}=0$ for the
first move). After the move to prime $b$, these differences telescope and the register
holds $\sum_j C_b(j,e_j)\bmod 2^{L}=\sum_j d_{b,j}(e_j)\bmod 2^{L}$, because
$L\le32$ ensures $2^{L}\mid2^{32}$. The final move to row $|P|$ returns every
register to $0$.

\paragraph{\ref{app:proofs}2. Trimmed-width bound.}
The true sum satisfies $V_b=\sum_j d_{b,j}(e_j)\le W_1(p_b-2)<W_1(2^{\ell}-1)<2^{L}$
for $L$ equal to the bit length of $W_1(2^{\ell}-1)$. The stored sum is
therefore the ordinary integer $V_b$,
and $L-\ell$ long-division steps reduce $V_b\bmod(p_b-1)$ exactly when
$W_1\le 2^{L-\ell}$.

\paragraph{\ref{app:proofs}3. Lazy-residue congruence, overflow bound, and canonicalization.}
Write $M_{j,a_j}=g_p^{\,a_j\,2^{w_{3a}j}}\bmod p$ and let $v_u$ be the $u$-th
$w_{3b}$-bit window of the $R$-bit register $r_{j-1}$. The lazy tables hold
$\mathsf{Tab}_j[u](a_j,v_u)=M_{j,a_j}\,v_u\,2^{w_{3b}u}\bmod p\in[0,p)$, and step $j$
computes
\[r_j=\sum_{u=0}^{W_3-1}\mathsf{Tab}_j[u](a_j,v_u)\equiv M_{j,a_j}r_{j-1}\pmod p.\]
When $W_3=\lceil R/w_{3b}\rceil\le 2^{R-\ell}$, $r_j<W_3 p\le 2^R$, so no
addition wraps modulo $2^R$. Before the unchanged CRT stage (Loop~4), a
canonicalization step of $R-\ell$ division steps reduces the final lazy residue $r_{J-1}$ to $r_{J-1}\bmod p$.

\paragraph{\ref{app:proofs}4. Reversing a checkpointed step.}
The retained predecessor $r_{j-1}$ supplies the same source digits $v_u$
when step $j$ is reversed. Subtracting the forward table entries therefore
returns the target to zero, including when its lazy value was greater than
$p$. The backward sweep visits each predecessor while its inputs are live.

\paragraph{\ref{app:proofs}5. Measurement phases.}
For a table value $\mathsf{Tab}(a)$ written to a $b$-bit target, the
amplitude for an X-basis result $\mu$ is
$2^{-b/2}(-1)^{\mu\cdot\mathsf{Tab}(a)}$. Its magnitude is independent of
the address. The recorded result and the address specify a diagonal
correction with the same sign, so their product removes the phase.
When multiple targets share the address, their signs multiply and their
parities add modulo two. Consequently each complete measurement record
has conditional map $K_\mu=c_\mu V_{\rm ref}$, where $c_\mu$ is independent
of the input and $\sum_\mu|c_\mu|^2=1$. Discarding the record preserves
the reference channel on arbitrary coherent inputs.

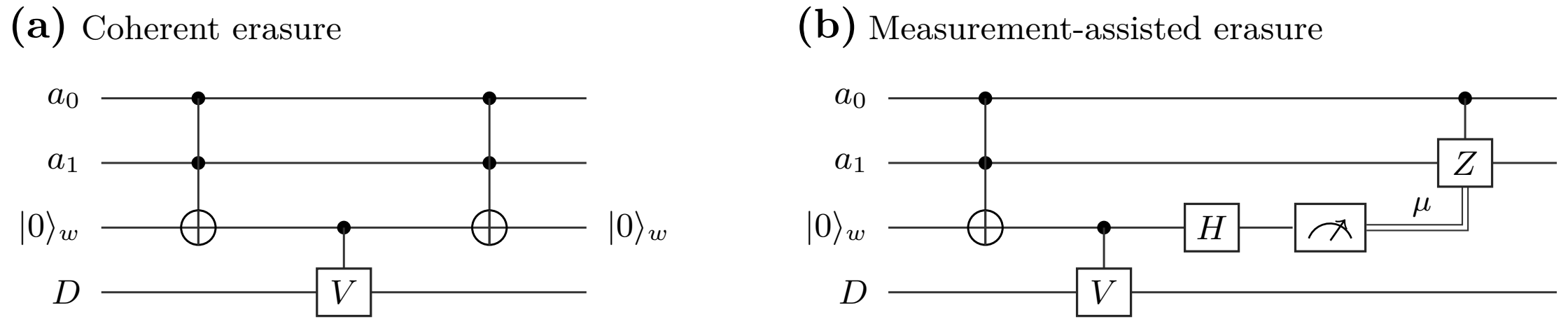
\begin{figure}[H]
\centering
\resizebox{0.92\linewidth}{!}{\input{tikz_phase}}
\caption{Two ways to erase a one-bit lookup after using its result.
Both circuits compute $a_0a_1$ into a clean work qubit and use it to
control $V$ on the data. (a) A second Toffoli returns the work qubit to
$\lvert0\rangle$. (b) A Hadamard followed by a computational-basis
measurement gives a classical bit $\mu$. The double line carries
$\mu$ to the phase correction on the address. The circuit applies a
controlled-$Z$ precisely when $\mu=1$. In either branch the address and
data have the same joint state as in (a).}
\label{fig:phase-example}
\end{figure}

Figure~\ref{fig:phase-example} gives an elementary example of this
phase correction~\cite{gidney2018and}. Let $a=(a_0,a_1)$ and
$t(a)=a_0a_1$. Before the lookup, write the joint address and data state
as $\sum_a \lvert a\rangle\lvert\phi_a\rangle$, with the amplitudes
absorbed into the data vectors. The computed bit controls an update
$V$ on the data. After an $X$-basis measurement of that bit, the
normalized joint state for outcome $\mu$ is
\[
  \sum_a (-1)^{\mu t(a)}\lvert a\rangle V^{t(a)}\lvert\phi_a\rangle .
\]
For $\mu=1$, a controlled-$Z$ on the two address qubits changes the sign
of precisely the terms with $a_0=a_1=1$. For $\mu=0$, no correction
is required. Both outcomes therefore leave the intended update on the
data, including when the data and address were initially entangled.

\paragraph{\ref{app:proofs}6. Direct writes.}
With a known zero target, adding a selected table entry and writing that
entry give the same basis value. The direct lookup used for the first
multiplication tabulates the composed function of the active windows.
These substitutions preserve the basis map, and the phase corrections of
\ref{app:proofs}5 preserve its coherent action. \hfill$\square$

\medskip
\begin{proposition}[From lane tails to a per-run error bound]\label{prop:mgf}
Let the $m$-bit exponent be uniformly superposed. For prime $p_i$ put
$h_i=(p_i-1)/2$. Window $j$ has $w_j=\min(w_1,m-jw_1)$ active bits and
contributes a table entry $d_{ij}(a)\in[0,h_i-1]$. Define
\[
 S_i=\sum_{j=0}^{W_1-1}d_{ij}(e_j),\qquad
 \mu_i=\sum_j2^{-w_j}\sum_{a=0}^{2^{w_j}-1}d_{ij}(a).
\]
Let $B_i$ be the integer offset actually used by the circuit. The $D$-bit
signed divider accepts $-c h_i\le S_i-B_i<c h_i$, where
$c=2^{D-(\ell-1)-1}$. This interval fits inside the signed register since
$h_i<2^{\ell-1}$. Write $a_i=B_i-c h_i$, $b_i=B_i+c h_i$.
Suppose $u_i$ bounds $\Pr[S_i<a_i \text{ or } S_i\ge b_i]$ for each
prime in $P$. If the compressed and full-width computations agree on the
complement of these events, their output states have trace distance at most
\[
 \eta_{\rm full}=\min\!\left\{1,\;2\sqrt{\min\{1,\sum_{i\in P}u_i\}}\right\}.
\]
The probability of any subsequent acceptance test changes by at most this
amount.
\end{proposition}

\begin{proof}
On a valid lane the signed division recovers the intended quotient parity
and reduced logarithm. The known offset is restored by the classical
multiplicative factor in the residue calculation. Let $\Pi$ project onto
the exponent strings for which any lane is invalid. The squared norm of
the projected input is the weight $q$ of this set, with
$q\leq\sum_{i\in P}u_i$ by the union bound. The two computations, with
measurement records retained coherently, agree on $(I-\Pi)$ by hypothesis.
Their output vectors therefore differ in norm by at most $2\sqrt q$.
Trace distance is at most this norm for pure states and cannot increase
when the measurement record is discarded. The union bound permits arbitrary
dependence between the prime failure events.
\end{proof}

\paragraph{Evaluating a single prime.}
Uniform exponent windows are independent, including a shortened last
window. For $\lambda>0$ their moment-generating functions give
\[
 \Pr[S_i\ge b_i]\le
 e^{-\lambda b_i}\prod_j
 \left(2^{-w_j}\sum_{a=0}^{2^{w_j}-1}e^{\lambda d_{ij}(a)}\right).
\]
Applying the same inequality to $-S_i$ with threshold $1-a_i$ bounds the
lower tail. The implementation rounds the exponential estimates outwards
using integer and rational arithmetic, so the sum of the two results
bounds the failure probability for that prime. A simpler option follows
from Hoeffding's inequality. If the distance from $\mu_i$ to both endpoints
is at least $t_i>0$, then $u_i=2\exp[-2t_i^2/(W_1(h_i-1)^2)]$ suffices.
Applying this expression to an entire prime set requires a valid margin
for each offset. The exact mean above depends on the table entries.

\paragraph{Tail adjustment used for the reported costs.}
For each of the 30 sampled-allowance parameter sets, the audit evaluates 27 prime indices spread across the ordered prime list (whereas the 12 gap-fill circuits evaluate exact integer MGF certificates across every prime in the set).
From the largest single-prime bound in that sample
$J$ it computes
\[
 \eta_{\rm audit}=2\sqrt{\min\{1,\;|P|\max_{i\in J}u_i\}}.
\]
The score deducts the recorded adjustment from $p_{\rm keep}$. As an
example, at $s=8$, $w_1=7$, $\ell=20$ and $D=25$, the largest sampled
bound is $3.68\times10^{-14}$. Extrapolation to $18{,}931$ primes gives
$5.28\times10^{-5}$, for which the score deducts $5.29\times10^{-5}$.
At $s=3$, $w_1=8$ and $D=26$, the corresponding sample gives
$1.68\times10^{-25}$ for $22{,}193$ primes. Both are extrapolations.
The exact bounds apply to the evaluated primes. An upper bound for the
entire computation also needs control of unexamined primes, either by
evaluating their tables or by proving a uniform bound for their offsets
and valid intervals. The reported factoring costs are conditional on this
sampling assumption.

\medskip
\begin{proposition}[A mask under a change of CRT coordinates]\label{prop:dyadic}
Let $K=2^f$, $M_t=\lfloor N/2^{n-f}\rfloor$ and $\alpha=K/M_t$, with
$1<\alpha<2$. The benchmark starts with a uniform mask on the consecutive
integers $0,\ldots,H-1$, where $H$ is a configured power of two and
$H<M_t$. The overlap of this mask with its unit translate is $(H-1)/H$.
If each of these values is first mapped to $\lfloor\alpha m\rfloor$ in
the dyadic register, the overlap with a unit translate there is
\[
 \frac{2(H-1)-\lfloor\alpha(H-1)\rfloor}{H}.
\]
\end{proposition}

\begin{proof}
The rescaled support is strictly increasing and starts at zero.
Its successive gaps have length one or two. The number of length-two
gaps is $\lfloor\alpha(H-1)\rfloor-(H-1)$ by summing the gaps. Each
remaining length-one gap contributes one point to the overlap, and
$H<M_t$ excludes a wraparound contribution. Normalising by $H$ gives
the stated expression.
\end{proof}

For the benchmark RSA-2048 modulus at $f=33$, $M_t=6{,}697{,}130{,}461$.
With a $26$-bit mask the two overlaps are approximately
$0.999999985$ and $0.717371$. This calculation concerns the indicated
rescaling of the benchmark mask. Li et al.~\cite{li2026} define a dyadic
CRT accumulator with its own rounding convention. In the experiments
reported here the benchmark subtracts the original mask modulo $M_t$.
Fusion retains this modulus throughout.

\section{Benchmark details}
\label{app:bench}

\paragraph{What a complete instance checks.}
The benchmark tracks branch values, relative phases, live registers and
primitive costs. It rejects unpriced non-Clifford operations, classical
reads of superposed values, overlapping operands, invalid mask preparations
and operations for which a runtime cost is missing.
The arithmetic runs over the entire prime set of each correctness instance.
There are 64 simulated branches at 32 bits, 32 at 64 bits, 16 at 128 bits
and eight for each of two settings on the same 330-bit RSA-100 modulus.
The second RSA-100 setting exercises the window and accumulator choices
used for the larger calculation. The verifier subtracts the recorded
mask modulo $M_t$ before comparing each branch with exact classical
modular exponentiation. It also checks the surviving exponent, work
registers and phases.

\paragraph{Projecting to RSA-2048.}
For the challenge modulus and base $3$, \path{rsabench/instances.py} (\texttt{stratified\_indices}) draws stratified \emph{individual} prime indices from the sorted prime list and packs them into lookup groups of size $k$, rather than sampling intact pre-formed groups from the full circuit; successive-prime differences and skipped zero coefficients are evaluated within those packed samples. We estimate the contribution of a
complete group without charging the final, partly filled group as if it
were full. Let $T(\nu)$ be the score of a circuit with $\nu$ sampled
primes. The first two sample sizes $\nu_1$ and $\nu_2$ round 24 and 120
up to multiples of $k$. We measure a third sample at
$\nu_3=\nu_1+(|P|\bmod k)$. The estimate is
\[
 T_{\rm shot}=T(\nu_3)+
 \frac{|P|-\nu_3}{\nu_2-\nu_1}\bigl[T(\nu_2)-T(\nu_1)\bigr].
\]
The anchor $T(\nu_3)$ includes the correct group remainder, and the slope
comes from adding only complete groups; the eight-replicate confidence intervals measure sampling variation under this estimator rather than full-circuit variance. A separate 330-bit comparison
checks this estimate against a run over every prime. Across all 42 primary
scores in the paper, every circuit passes the $1\,\%$ RSA-100 projection check (maximum primary discrepancy $0.469\,\%$; maximum eight-replicate mean discrepancy $0.53\,\%$ at 1,393 qubits), and the peak qubit counts agree.
On the earlier 30-circuit subset (240 further comparisons with fresh samples), eight exceeded $1\,\%$ in
absolute value, all at 1,379 or 1,393 logical qubits, with a largest
discrepancy of $1.99\,\%$.

\paragraph{Worked estimate at 1,685 logical qubits.}
For $s=8$, $w_1=7$, $\ell=20$ and $k=8$, the full list has 18,931
primes. Samples of 24 and 120 primes differ by twelve complete groups.
A 27-prime sample contains three complete groups and the same three-prime
remainder as the full circuit. The estimator adds
$(18{,}931-27)/(120-24)$ times the difference between the first two
counts to the 27-prime count. Applying this calculation to each primitive
also supplies the operation counts used by the architecture models.
The primary sample gives 188.90 million Toffolis per execution and an
adjusted kept fraction of 0.98621. Thus $9(0.18890)/0.98621=1.724$
billion Toffolis per factoring to the precision shown. Repeating the
sampling eight times produces the mean and interval in Table~\ref{tab:all}.
Every primary estimate lies within 0.24\,\% of its eight-replicate mean.

\paragraph{Harness for other $(s,f,k)$.}
\path{rsabench/score.py} fixes $s=8$, $f=33$ and $k\le16$. The other
settings use \path{sweep/eval_extended.py}, which calls the same primitive
costs, circuit runner, strict checks and instance code. The underlying
scores agree in the common parameter range. The saved RSA-2048 data
include a separate audit of 37 prime-system files for arithmetic capacity
and forbidden zero residues among window multipliers. These checks
establish the suitability of the prime system for constructing tables.
The subsequent tail calculation samples tables as specified in
Appendix~\ref{app:proofs}.

\paragraph{Serial reference.}
For a candidate with peak width $Q$, the comparator takes the least
predicted cost among serial settings whose width is at most $Q$.
The search covers Gidney's \frontierGrid{}-setting grid at $k=1$.
The serial model receives the benefits of measured cleanup, sliced
non-restoring division, banked and direct lookups, the half-modulus
quotient, compressed lanes, fusion of loops~3 and~4, reflected CRT digits
and phase lookups that skip zero coefficients. Its qubit convention omits
some carry and bookkeeping qubits, which favours the comparator. The
smallest feasible serial setting in this grid has 1,291 logical qubits.

\section{Primitive checks}
\label{app:refinements}

Small instances allow exhaustive checks of arithmetic identities.
Larger register widths and complete benchmark runs exercise code paths
used in the resource estimates. The recorded checks include the following.
\begin{itemize}
\item The CRT shortcut was checked over 82,366--98,638 representatives at
  prime widths $\ell\in\{19,20,21\}$. Tests of non-restoring division
  covered 21,080 inputs. The half-modulus quotient was checked on 14,460
  small and 1,080 full-width signed branches. Further tests exercised
  overlapping nine-prime words, banked lookups, narrow adders, and
  folded and reflected CRT digits.
\item Ten tests in \path{arith/test_primitives.py} exercised the signed
  divider and its reversal, centred arithmetic, recovery of the CRT
  weight, stage fusion, monomial lookups, bilinear traversal, reflected
  residue arithmetic, split phase trees and two-ancilla addition.
\item Four tests in \path{arch/test_arch_primitives.py} covered batched
  rolling lookups up to $k=37$, the two-chunk CRT update, chunked
  arithmetic, and cleanup of work registers and phases.
\item The tail calculation in \path{audit/certify_dlog_tail.py} evaluates
  the per-prime inequalities in Proposition~\ref{prop:mgf} using integers
  and rational numbers. Its inputs include all active entries of each
  window table for the sampled prime.
\end{itemize}

\section{All circuits and further results}
\label{app:validation}

\begin{table}[!htbp]
\centering\small
\resizebox{\linewidth}{!}{% Generated by audit/build_paper_assets.py
\begin{tabular}{rrrrrrrrrrrr}
\toprule
$Q$ & $s$ & $k$ & $\ell$ & $w_1$ & $f$ & $D$ & \shortstack{$T_{\rm fact}/10^9$\\(mean $\pm$ 95\,\% CI)} & \shortstack{330-bit\\cross-check} & $\eta_{\rm audit}$ & \shortstack{Serial ref.\\$/10^9$} & Change\\
\midrule
1,379 & 8 & 1 & 19 & 8 & 33 & $24\mathrm{s}$ & $7.1000\pm0.0345$ & $+0.06\pm1.07$\,\% & $1.5\times10^{-6}$ & 4.477 & $+58.6$\,\%\\
1,393 & 8 & 1 & 19 & 8 & 33 & $24\mathrm{s}$ & $6.3037\pm0.0103$ & $+0.53\pm0.76$\,\% & $1.5\times10^{-6}$ & 4.467 & $+41.1$\,\%\\
1,403 & 8 & 1 & 20 & 7 & 33 & $25\mathrm{s}$ & $4.2514\pm0.0031$ & $+0.06\pm0.15$\,\% & $5.3\times10^{-5}$ & 4.227 & $+0.6$\,\%\\
1,450 & 8 & 2 & 20 & 7 & 33 & $25\mathrm{s}$ & $2.9896\pm0.0011$ & $+0.05\pm0.16$\,\% & $5.3\times10^{-5}$ & 3.936 & $-24.0$\,\%\\
1,459 & 8 & 2 & 20 & 7 & 33 & $25\mathrm{s}$ & $2.9677\pm0.0007$ & $-0.09\pm0.14$\,\% & $5.3\times10^{-5}$ & 3.936 & $-24.6$\,\%\\
1,550 & 8 & 5 & 20 & 7 & 33 & $25\mathrm{s}$ & $2.2276\pm0.0015$ & $+0.02\pm0.24$\,\% & $5.3\times10^{-5}$ & 3.548 & $-37.2$\,\%\\
1,597 & 8 & 5 & 20 & 7 & 33 & $25\mathrm{s}$ & $2.0352\pm0.0009$ & $-0.06\pm0.15$\,\% & $5.3\times10^{-5}$ & 3.372 & $-39.6$\,\%\\
1,649 & 8 & 6 & 20 & 7 & 33 & $25\mathrm{s}$ & $1.8626\pm0.0006$ & $-0.03\pm0.18$\,\% & $5.3\times10^{-5}$ & 3.297 & $-43.5$\,\%\\
1,685 & 8 & 8 & 20 & 7 & 33 & $25\mathrm{s}$ & $1.7244\pm0.0007$ & $-0.03\pm0.09$\,\% & $5.3\times10^{-5}$ & 3.203 & $-46.2$\,\%\\
1,796 & 8 & 12 & 20 & 7 & 33 & $25\mathrm{s}$ & $1.5559\pm0.0009$ & $+0.03\pm0.20$\,\% & $5.3\times10^{-5}$ & 3.078 & $-49.5$\,\%\\
1,995 & 8 & 16 & 20 & 7 & 33 & $25\mathrm{s}$ & $1.4565\pm0.0007$ & $-0.08\pm0.26$\,\% & $5.3\times10^{-5}$ & 2.958 & $-50.8$\,\%\\
2,025 & 8 & 16 & 20 & 8 & 33 & $25\mathrm{s}$ & $1.3535\pm0.0005$ & $-0.07\pm0.18$\,\% & $1.2\times10^{-6}$ & 2.958 & $-54.2$\,\%\\
2,050 & 8 & 16 & 20 & 8 & 33 & $25\mathrm{s}$ & $1.3479\pm0.0009$ & $-0.04\pm0.12$\,\% & $1.2\times10^{-6}$ & 2.958 & $-54.4$\,\%\\
\midrule
1,271 & 14 & 1 & 19 & 7 & 33 & $24\mathrm{s}$ & $11.0210\pm0.0019$ & $+0.01\pm0.04$\,\% & $4.1\times10^{-6}$ & --- & ---\\
1,291 & 14 & 1 & 19 & 7 & 33 & $24\mathrm{s}$ & $6.3640\pm0.0071$ & $-0.01\pm0.11$\,\% & $4.1\times10^{-6}$ & 6.145 & $+3.6$\,\%\\
1,298 & 14 & 1 & 20 & 7 & 33 & $25\mathrm{s}$ & $5.9880\pm0.0024$ & $+0.13\pm0.12$\,\% & $5.2\times10^{-6}$ & 6.013 & $-0.4$\,\%\\
1,342 & 14 & 2 & 20 & 7 & 33 & $25\mathrm{s}$ & $4.2688\pm0.0018$ & $-0.03\pm0.16$\,\% & $5.2\times10^{-6}$ & 5.093 & $-16.2$\,\%\\
1,378 & 14 & 3 & 20 & 6 & 33 & 25 & $3.7506\pm0.0016$ & $-0.07\pm0.15$\,\% & $7.3\times10^{-5}$ & 4.487 & $-16.4$\,\%\\
1,400 & 10 & 2 & 20 & 7 & 33 & $25\mathrm{s}$ & $3.4030\pm0.0017$ & $+0.07\pm0.16$\,\% & $9.9\times10^{-6}$ & 4.268 & $-20.3$\,\%\\
1,500 & 10 & 5 & 20 & 7 & 33 & $25\mathrm{s}$ & $2.5435\pm0.0015$ & $-0.08\pm0.35$\,\% & $9.9\times10^{-6}$ & 3.669 & $-30.7$\,\%\\
1,544 & 7 & 4 & 20 & 7 & 33 & 25 & $2.2797\pm0.0011$ & $-0.07\pm0.17$\,\% & $3.0\times10^{-5}$ & 3.548 & $-35.7$\,\%\\
1,703 & 7 & 7 & 20 & 7 & 37 & 25 & $1.6627\pm0.0008$ & $-0.02\pm0.16$\,\% & $3.0\times10^{-5}$ & 3.125 & $-46.8$\,\%\\
1,750 & 6 & 7 & 20 & 7 & 33 & 25 & $1.5517\pm0.0007$ & $+0.11\pm0.15$\,\% & $6.0\times10^{-5}$ & 3.118 & $-50.2$\,\%\\
1,751 & 7 & 9 & 20 & 7 & 35 & 25 & $1.5466\pm0.0010$ & $+0.06\pm0.16$\,\% & $3.0\times10^{-5}$ & 3.118 & $-50.4$\,\%\\
1,861 & 5 & 8 & 20 & 7 & 35 & 25 & $1.3799\pm0.0005$ & $-0.04\pm0.16$\,\% & $1.1\times10^{-4}$ & 3.047 & $-54.7$\,\%\\
1,893 & 5 & 10 & 20 & 7 & 35 & 25 & $1.3028\pm0.0006$ & $-0.04\pm0.16$\,\% & $1.1\times10^{-4}$ & 3.014 & $-56.8$\,\%\\
1,948 & 5 & 12 & 20 & 7 & 33 & $25\mathrm{s}$ & $1.2665\pm0.0006$ & $+0.02\pm0.09$\,\% & $1.8\times10^{-4}$ & 3.014 & $-58.0$\,\%\\
2,050 & 4 & 12 & 20 & 7 & 33 & $25\mathrm{s}$ & $1.1930\pm0.0003$ & $+0.04\pm0.06$\,\% & $6.3\times10^{-4}$ & 2.958 & $-59.7$\,\%\\
2,123 & 4 & 15 & 20 & 7 & 33 & 25 & $1.1302\pm0.0004$ & $-0.02\pm0.11$\,\% & $3.5\times10^{-4}$ & 2.958 & $-61.8$\,\%\\
2,174 & 5 & 16 & 20 & 8 & 33 & 25 & $1.1105\pm0.0005$ & $+0.02\pm0.21$\,\% & $1.0\times10^{-5}$ & 2.958 & $-62.5$\,\%\\
2,185 & 5 & 15 & 20 & 8 & 37 & $25\mathrm{s}$ & $1.1222\pm0.0006$ & $-0.00\pm0.16$\,\% & $2.0\times10^{-4}$ & 2.958 & $-62.1$\,\%\\
2,226 & 4 & 14 & 20 & 8 & 35 & 25 & $1.0775\pm0.0008$ & $+0.09\pm0.20$\,\% & $3.2\times10^{-5}$ & 2.958 & $-63.6$\,\%\\
2,276 & 4 & 16 & 20 & 8 & 33 & 25 & $1.0375\pm0.0007$ & $-0.01\pm0.19$\,\% & $3.2\times10^{-5}$ & 2.958 & $-64.9$\,\%\\
2,298 & 4 & 16 & 20 & 8 & 37 & $25\mathrm{s}$ & $1.0288\pm0.0005$ & $-0.05\pm0.12$\,\% & $5.8\times10^{-5}$ & 2.958 & $-65.2$\,\%\\
2,358 & 4 & 19 & 20 & 8 & 37 & $25\mathrm{s}$ & $0.9805\pm0.0004$ & $+0.05\pm0.14$\,\% & $5.8\times10^{-5}$ & 2.932 & $-66.6$\,\%\\
2,432 & 4 & 22 & 20 & 8 & 33 & 25 & $0.9471\pm0.0005$ & $+0.00\pm0.16$\,\% & $3.2\times10^{-5}$ & 2.932 & $-67.7$\,\%\\
2,554 & 4 & 25 & 20 & 8 & 37 & $25\mathrm{s}$ & $0.9224\pm0.0003$ & $-0.03\pm0.14$\,\% & $5.8\times10^{-5}$ & 2.932 & $-68.5$\,\%\\
2,649 & 4 & 30 & 20 & 8 & 37 & $25\mathrm{s}$ & $0.8892\pm0.0004$ & $-0.00\pm0.21$\,\% & $5.8\times10^{-5}$ & 2.932 & $-69.7$\,\%\\
2,925 & 3 & 34 & 20 & 8 & 37 & $25\mathrm{s}$ & $0.8404\pm0.0006$ & $+0.13\pm0.21$\,\% & $3.4\times10^{-4}$ & 2.932 & $-71.3$\,\%\\
2,977 & 3 & 36 & 20 & 8 & 37 & $25\mathrm{s}$ & $0.8310\pm0.0003$ & $+0.05\pm0.13$\,\% & $3.4\times10^{-4}$ & 2.932 & $-71.7$\,\%\\
2,994 & 3 & 34 & 20 & 8 & 37 & $26\mathrm{s}$ & $0.8623\pm0.0003$ & $-0.00\pm0.16$\,\% & $1.7\times10^{-25}$ & 2.932 & $-70.6$\,\%\\
2,998 & 3 & 37 & 20 & 8 & 37 & $25\mathrm{s}$ & $0.8277\pm0.0004$ & $+0.03\pm0.13$\,\% & $3.4\times10^{-4}$ & 2.932 & $-71.8$\,\%\\
\bottomrule
\end{tabular}
}
\caption{Logical costs for 13 circuits at $s=8$ (upper block) and 29 at
other repetition settings (lower block). Both $T_{\rm fact}$ and the
signed 330-bit projection error give the mean and 95\,\% confidence
half-width across eight replicates (336 replicates across all 42 circuits).
In column $D$, a trailing ``s'' ($24\mathrm{s}, 25\mathrm{s}, 26\mathrm{s}$) marks the 30 circuits whose compression allowance $\eta_{\rm audit}$ extrapolates from 27 sampled primes per $(s,\ell,w_1)$ set, whereas an unsuffixed $25$ marks the 12 gap-fill circuits certified over all primes of the $(s,\ell,w_1)$ prime list via exact integer MGF certificates under the centred half-period rule (wraparound gap $37$, except $33$ at $1,378$ qubits; lane width $D=25$).
The final column
compares the mean cost with the serial model at the same qubit cap.
A dash denotes a cap below the minimum width in the serial grid.}
\label{tab:all}
\end{table}

\subsection{Controlled arithmetic comparisons}
\label{app:ablation}

A difference between two points on the frontier generally combines several
decisions. For example, increasing a lookup window can change the number of
window additions as well as the number of auxiliary primes. To isolate
individual mechanisms, Table~\ref{tab:ablations} holds the other listed
parameters fixed. These comparisons use $\ell=21$ and expose the effect of
a single change even when the resulting circuit would not be selected for
the RSA-2048 frontier.

\input{ablation_table}

At $w_1=6$ and $k=1$, storing four checkpoints reduces the factoring cost
from 5.688 to 5.055 billion Toffolis. This is an 11.1\,\% reduction for 88
extra live qubits. The work saved is the inverse modular arithmetic that
would reconstruct the discarded residues. Once those checkpoints exist,
lazy arithmetic gives a further 3.9\,\% reduction, to 4.856 billion, at a
cost of ten more peak qubits. Changing instead from one to three primes
per lookup while keeping all four checkpoints and canonical residues
takes 5.055 billion down to 3.502 billion, a 30.7\,\% reduction. Sharing
has a large effect because address decoding accounts for so much of the
starting circuit.

At $w_1=8$ and $k=6$, switching to lazy residues lowers 2.756 billion to
2.605 billion, or 5.5\,\%. Both runs peak at 1,690 qubits. The wider residue
can fit into space left available at that stage of the computation. This
example illustrates two reasons to avoid adding percentages from the
different rows. The starting costs differ, and the saving from removing
reductions depends on how much work remains in the residue stage.

\paragraph{Core transformations at $k=1$.}
Table~\ref{tab:ladder} holds the window and prime sizes fixed while
adding the mechanisms in sequence. Trimming changes the width of a
single lane. Keeping its intermediate residues then trades extra live
qubits for cheaper reverse arithmetic. Lazy accumulation changes the
cost of that checkpointed computation.

\begin{table}[!htbp]
\centering\small
\begin{tabular}{lrrrr}
\toprule
Circuit ($k=1$, $w_1=6$) & $Q$ & $T_{\rm shot}$ & \shortstack{Change in\\$T_{\rm shot}$ (\%)} & \shortstack{Change in\\$Q T_{\rm shot}$ (\%)}\\
\midrule
Published configuration & \baseQ & $\baseTshot$ & --- & --- \\
+ trimmed register & \trimOnlyQ & $\trimOnlyTshot$ & \trimOnlydT & \trimOnlydQT \\
+ all checkpoints & \trimCkptQ & $\trimCkptTshot$ & \trimCkptdT & \trimCkptdQT \\
+ lazy residues & \lazyOnlyQ & $\lazyOnlyTshot$ & \lazyOnlydT & \lazyOnlydQT \\
\bottomrule
\end{tabular}
\caption{Cumulative changes at $k=1$, $s=8$, $\ell=21$ and $w_1=6$.
Both percentage columns use the first row as their denominator.
$T_{\rm shot}$ counts Toffolis in a single run.}
\label{tab:ladder}
\end{table}

\paragraph{Phase-table pricing.}
Explicit-gate checks over 36 small chains with moduli 3, 5, 7 and 13 and
lengths one to three confirmed the binary phase coefficients and uniform
measurement transcripts.
Replacing the benchmark's $\lfloor\sqrt{2^a}\rfloor$ phase-table price by the
constructive half-split bound
$C_\phi(a)=2^{\lceil a/2\rceil}+2^{\lfloor a/2\rfloor}-a-2$ raises the totals of
the 30 primary scores by at most \phaseMaxChange\,\%. For example, the
2,358-qubit circuit moves from $0.9802$ to $0.9971$ billion projected
Toffolis under this substitution. Those numbers use the same tail
allowance and primary sample. The LDPC-cat and Pinnacle projections use
$C_\phi(a)$ directly.

\begin{figure}[!htbp]
\centering
\includegraphics[width=\linewidth]{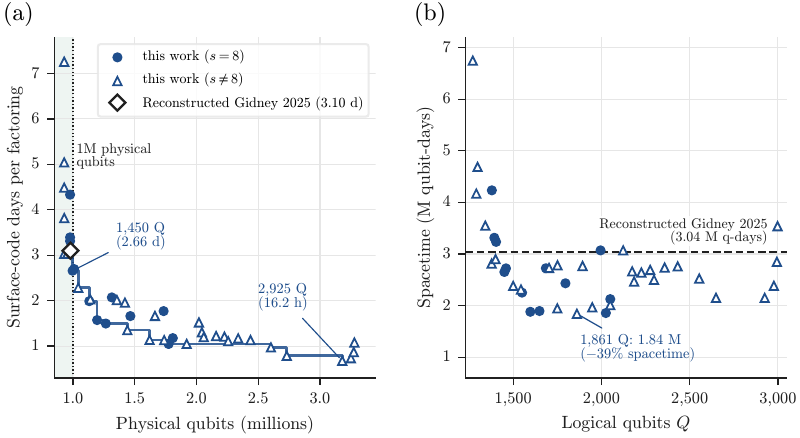}
\caption{Surface-code runtime and spacetime volume for the 42 circuits.
(a) Physical qubits versus days per factoring, with the 1M physical-qubit
boundary shaded and Gidney's 2025 baseline marked by the open diamond.
(b) Spacetime volume (physical qubits $\times$ days) versus logical width
$Q$, showing the 39\,\% reduction below Gidney's 2025 baseline at 1,861
logical qubits.}
\label{fig:landscape}
\end{figure}

\paragraph{Physical spacetime.}
Figure~\ref{fig:landscape} places duration and physical capacity on
separate axes. The 1,861-logical-qubit circuit minimises their product
($1.84$ million qubit-days) among the 42 designs, while the
2,925-logical-qubit circuit finishes sooner by using a larger processor.

\FloatBarrier
\section{Hardware models and projections}
\label{app:transfer}

\subsection{Surface-code timing and footprint}

The surface-code model prices each recorded addition, comparison, value
lookup and phase lookup using the durations in Gidney's cost
model~\cite{gidney2025}. Its code cycle is $1\,\mu$s and its measurement
reaction time is $10\,\mu$s. The runtime for a factoring is the sum of
the primitive durations for a shot, multiplied by the expected number of
shots. For the footprint, the $m$ exponent qubits occupy compact storage.
The reference layout reserves $17\times18=306$ active patches for its
working area, factories and routing. We estimate the physical-qubit count as
\[
  440m+1{,}352\bigl[306+\max(0,Q-1{,}435)\bigr].
\]
Thus the reference working area is retained below 1,435 logical qubits,
and each additional logical qubit above that peak costs one active patch.
The layout estimate charges a full patch for each additional qubit and
holds the routing allowance fixed.

For $s=8$ the step from 1,450 to 1,459 logical qubits adds nine active
patches, or 12,168 physical qubits. The latter circuit needs 1,009,360
physical qubits and takes 2.70 days. It crosses the one-million-qubit
boundary while slightly increasing the time. Across all 42 circuits,
the minimum spacetime volume under this timing model is 1.8425 million physical-qubit days at 1,861 logical qubits
($1.621$ million physical qubits for $1.137$ days), compared with
approximately $3.04$ million physical-qubit days at the reference setting; within the $s=8$ subset, the minimum is 1.8566 million physical-qubit days at 2,025 logical qubits ($1.775$ million physical qubits for $1.046$ days), about $0.8\,\%$ higher, so the ordering between these two points is specific to the timing model.

\subsection{Neutral-atom processor and memory}

Both neutral-atom configurations use the LP24 memory of Cain
et al.~\cite{cain2026}. A surgery operation takes
$\tau_s=(2d_p/3)\tau_{\rm code}$ with $\tau_{\rm code}=1$\,ms and
processor distances $d_p=20$ (balanced) and $18$ (space-efficient).
On the 148-qubit balanced processor, a lookup with $a$ address bits
can return up to $148-2a$ word bits in one piece. A wider word is split
as in Appendix~E.3 of Ref.~\cite{cain2026}, repeating the address work
for each piece. The space-efficient processor has seven word qubits.
The reconstruction prices a lookup of $N>2$ entries and $b$ output bits
at $(15bN/7)\tau_s$. It prices comparisons as additions and phase
lookups with more than four entries as reads with a one-bit output.

For additions and lookups together, the reconstructed serial workload
requires about 10 surgery cycles per Toffoli on the balanced processor
and about 52 on the space-efficient processor. The corresponding
published estimates are about 10 and 43. The discrepancy on the
space-efficient processor limits the precision of its absolute years.
The references in Table~\ref{tab:hw} use the fastest evaluated serial
window on each processor, $w_1=7$ and $5$, respectively.

We check memory capacity separately by comparing the circuit's full
peak $Q$ to 1,480. At 1,450 qubits the balanced and space-efficient
times are 29.3 and 96.2 years. With the timing parameters held fixed, a
larger memory gives 6.9 years on the balanced processor at 2,925 qubits
and 14.4 years on the space-efficient processor at 2,994 qubits. Each
estimate would require a physical design for the additional memory,
including its transport latency and logical failure rate.

\subsection{LDPC-cat work region and residual overhead}

The serial reconstruction of Pappalardo et al.~\cite{pappalardo2026}
gives 22,593 physical qubits, 7.841 hours per shot and 3.137 days per
factoring. For each candidate we allocate $Q+1+\delta_\phi$ work
slots. The additional slot serves the adder, and $\delta_\phi$ covers
the maximum extra workspace for the constructive phase-table
implementation. We retain the published code distances, resize the work
and buffer regions, and recalculate shot survival using the new
allocation, shot length and resource-state counts.

The serial schedule spends $H_0=2.876\times10^9$ cycles per shot on
residual operations such as movement, preparation and measurement.
A compiled schedule for the candidate would determine its own residual
cost $H$. Table~\ref{tab:hw} evaluates the scenarios $H=H_0$ and $H=0$,
charging the recorded primitives in both. At 2,925 logical qubits,
removing the residual cost changes 0.561 days to 0.412 days. For a
matched-footprint comparison we select the fastest listed single-chip
configuration in Table~XXXV of Ref.~\cite{pappalardo2026} that fits
the candidate's physical budget. This credits parallel engines whenever
they fit.

\subsection{Pinnacle~1.0 blocks and survival}

For Pinnacle~1.0~\cite{pinnacle2026} (arXiv:2602.11457v1; distinct from the later Pinnacle~1.2 release~\cite{pinnacle12}), we use one processing unit, a
batchwise accumulator, physical error parameter $p=10^{-3}$, code
distance 24 and a $26\,\mu$s logical cycle. For the four target runtimes
in Table~VI of Ref.~\cite{pinnacle2026}, the reconstructed physical-qubit
counts agree with the rounded published counts to within $0.5\,\%$.
The candidate occupies
$\lceil m/16\rceil$ memory blocks and
$\lceil(Q-m+1+\delta_\phi)/16\rceil$ processing blocks. For a shot
lasting $\mathcal T$ logical cycles at allocated logical capacity $N_L$,
we use
\[
 p_S=(1-p_L)^{N_L\mathcal T}(1-p_T)^{4A}.
\]
Here $p_L$ is the failure probability per logical qubit-cycle, $p_T$
is the error probability per consumed T state, and $A$ is the model's
Toffoli-equivalent count per shot. The second factor prices four T
states per Toffoli equivalent. A larger unit therefore changes both the
available work space and the number of qubit-cycles exposed to failure.
The transfer assumes the published timing and error rates continue to
apply as the unit grows.

For each row we retune the reference's parallelism within the candidate's
physical footprint. At 1,271 logical qubits the candidate has a time
estimate of 22.50 days, but no searched reference fits that budget.

\begin{table}[!htbp]
\centering\small
\resizebox{\linewidth}{!}{% Generated from audited hardware projections
\begin{tabular}{rrrrrrrrrrrrr}
\toprule
$Q$ & $s$ & $k$ & \shortstack{Surface code\\(days)} & \shortstack{Physical\\qubits ($10^6$)} & \shortstack{Balanced proc.\\atoms (y)} & \shortstack{Space-eff. proc.\\atoms (y)} & \shortstack{LDPC-cat\\$H_0$ / $0$ (d)} & \shortstack{Cat phys.\\($10^3$)} & \shortstack{Cat ref.\\(d)} & \shortstack{Pinnacle~1.0\\(d)} & \shortstack{Pin. ref.\\(d)} & \shortstack{Pinnacle\\change}\\
\midrule
\multicolumn{3}{l}{Reference} & 3.10 & 0.98 & 25.4 & 109.5 & 3.14 & 22.6 & --- & --- & --- & ---\\
\midrule
1,379 & 8 & 1 & 4.33 & 0.977 & 40.9 & 148.6 & 3.04 / 2.71 & 22.8 & 3.14 & 14.56 & 21.99 & $-$33.8\,\%\\
1,393 & 8 & 1 & 3.39 & 0.977 & 29.8 & 151.0 & 2.69 / 2.36 & 22.8 & 3.14 & 12.86 & 20.78 & $-$38.1\,\%\\
1,403 & 8 & 1 & 3.31 & 0.977 & 34.8 & 91.7 & 2.18 / 1.86 & 23.9 & 3.14 & 8.68 & 18.45 & $-$52.9\,\%\\
1,450 & 8 & 2 & 2.66 & 0.997 & 29.3 & 96.2 & 1.72 / 1.39 & 27.2 & 3.14 & 6.05 & 14.99 & $-$59.6\,\%\\
1,459 & 8 & 2 & 2.70 & 1.009 & 29.7 & 93.1 & 1.73 / 1.40 & 27.3 & 3.14 & 6.00 & 14.50 & $-$58.6\,\%\\
1,550 & 8 & 5 & 1.99 & 1.132 & 21.3$^\dagger$ & 89.2$^\dagger$ & 1.34 / 1.02 & 28.9 & 3.14 & 4.47 & 9.70 & $-$53.9\,\%\\
1,597 & 8 & 5 & 1.57 & 1.196 & 15.9$^\dagger$ & 90.5$^\dagger$ & 1.21 / 0.89 & 29.7 & 1.59 & 4.11 & 9.03 & $-$54.5\,\%\\
1,649 & 8 & 6 & 1.50 & 1.266 & 15.5$^\dagger$ & 90.8$^\dagger$ & 1.16 / 0.84 & 30.6 & 1.59 & 3.75 & 7.89 & $-$52.5\,\%\\
1,685 & 8 & 8 & 2.07 & 1.315 & 24.1$^\dagger$ & 66.0$^\dagger$ & 1.17 / 0.84 & 31.3 & 1.59 & 3.47 & 7.35 & $-$52.8\,\%\\
1,796 & 8 & 12 & 1.66 & 1.465 & 17.5$^\dagger$ & 50.4$^\dagger$ & 1.09 / 0.76 & 36.0 & 1.06 & 3.16 & 5.27 & $-$40.0\,\%\\
1,995 & 8 & 16 & 1.77 & 1.734 & 20.5$^\dagger$ & 39.7$^\dagger$ & 1.06 / 0.73 & 35.4 & 1.06 & 2.93 & 4.28 & $-$31.4\,\%\\
2,025 & 8 & 16 & 1.05 & 1.775 & 11.2$^\dagger$ & 94.4$^\dagger$ & 0.94 / 0.61 & 36.0 & 1.06 & 2.72 & 4.21 & $-$35.4\,\%\\
2,050 & 8 & 16 & 1.18 & 1.808 & 12.8$^\dagger$ & 97.5$^\dagger$ & 0.95 / 0.63 & 37.7 & 1.06 & 2.71 & 3.93 & $-$31.1\,\%\\
\midrule
1,271 & 14 & 1 & 7.26 & 0.929 & 71.3 & 185.3 & 4.87 / 4.32 & 22.8 & 3.14 & 22.50 & --- & ---\\
1,291 & 14 & 1 & 4.49 & 0.929 & 45.7 & 139.8 & 3.20 / 2.66 & 22.8 & 3.14 & 12.95 & 28.71 & $-$54.9\,\%\\
1,298 & 14 & 1 & 5.04 & 0.929 & 54.0 & 124.1 & 3.23 / 2.69 & 23.9 & 3.14 & 12.18 & 23.19 & $-$47.5\,\%\\
1,342 & 14 & 2 & 3.82 & 0.929 & 42.2 & 136.9 & 2.51 / 1.97 & 26.0 & 3.14 & 8.61 & 17.14 & $-$49.8\,\%\\
1,378 & 14 & 3 & 3.03 & 0.929 & 31.3 & 105.6 & 2.24 / 1.70 & 26.0 & 3.14 & 7.55 & 15.54 & $-$51.4\,\%\\
1,400 & 10 & 2 & 3.04 & 0.955 & 33.5 & 109.4 & 1.98 / 1.58 & 26.3 & 3.14 & 6.88 & 16.05 & $-$57.1\,\%\\
1,500 & 10 & 5 & 2.28 & 1.043 & 24.5$^\dagger$ & 101.3$^\dagger$ & 1.55 / 1.16 & 28.0 & 3.14 & 5.11 & 10.97 & $-$53.5\,\%\\
1,544 & 7 & 4 & 2.03 & 1.141 & 21.8$^\dagger$ & 83.2$^\dagger$ & 1.31 / 1.03 & 28.8 & 3.14 & 4.59 & 10.31 & $-$55.5\,\%\\
1,703 & 7 & 7 & 2.01 & 1.356 & 23.4$^\dagger$ & 68.5$^\dagger$ & 1.09 / 0.81 & 31.6 & 1.59 & 3.36 & 6.88 & $-$51.2\,\%\\
1,750 & 6 & 7 & 1.35 & 1.441 & 14.0$^\dagger$ & 69.1$^\dagger$ & 0.97 / 0.72 & 32.4 & 1.59 & 3.14 & 6.66 & $-$52.9\,\%\\
1,751 & 7 & 9 & 1.96 & 1.421 & 22.7$^\dagger$ & 56.9$^\dagger$ & 1.05 / 0.77 & 32.4 & 1.59 & 3.12 & 6.66 & $-$53.1\,\%\\
1,861 & 5 & 8 & 1.14 & 1.621 & 11.7$^\dagger$ & 57.9$^\dagger$ & 0.86 / 0.64 & 34.3 & 1.59 & 2.78 & 5.64 & $-$50.7\,\%\\
1,893 & 5 & 10 & 1.66 & 1.664 & 19.3$^\dagger$ & 45.8$^\dagger$ & 0.87 / 0.65 & 34.9 & 1.59 & 2.63 & 5.27 & $-$50.1\,\%\\
1,948 & 5 & 12 & 1.13 & 1.738 & 11.7$^\dagger$ & 44.4$^\dagger$ & 0.83 / 0.61 & 35.9 & 1.06 & 2.55 & 4.88 & $-$47.6\,\%\\
2,050 & 4 & 12 & 1.05 & 1.921 & 10.7$^\dagger$ & 42.0$^\dagger$ & 0.76 / 0.58 & 37.7 & 1.06 & 2.41 & 4.56 & $-$47.2\,\%\\
2,123 & 4 & 15 & 1.52 & 2.020 & 17.7$^\dagger$ & 33.0$^\dagger$ & 0.77 / 0.59 & 38.9 & 1.06 & 2.28 & 4.13 & $-$44.9\,\%\\
2,174 & 5 & 16 & 1.30 & 2.044 & 15.5$^\dagger$ & 76.2$^\dagger$ & 0.76 / 0.54 & 39.8 & 1.06 & 2.23 & 3.64 & $-$38.8\,\%\\
2,185 & 5 & 15 & 1.20 & 2.059 & 14.2$^\dagger$ & 79.1$^\dagger$ & 0.75 / 0.53 & 38.8 & 1.06 & 2.26 & 3.64 & $-$38.0\,\%\\
2,226 & 4 & 14 & 1.22 & 2.159 & 14.6$^\dagger$ & 80.5$^\dagger$ & 0.71 / 0.52 & 40.8 & 1.06 & 2.16 & 3.41 & $-$36.5\,\%\\
2,276 & 4 & 16 & 1.21 & 2.227 & 14.4$^\dagger$ & 72.1$^\dagger$ & 0.69 / 0.51 & 41.6 & 1.06 & 2.08 & 3.41 & $-$38.9\,\%\\
2,298 & 4 & 16 & 1.11 & 2.256 & 13.2$^\dagger$ & 70.8$^\dagger$ & 0.68 / 0.50 & 40.7 & 1.06 & 2.07 & 3.19 & $-$35.2\,\%\\
2,358 & 4 & 19 & 1.17 & 2.337 & 13.9$^\dagger$ & 59.6$^\dagger$ & 0.67 / 0.49 & 41.8 & 1.06 & 1.98 & 3.15 & $-$37.2\,\%\\
2,432 & 4 & 22 & 1.13 & 2.437 & 13.4$^\dagger$ & 52.6$^\dagger$ & 0.66 / 0.48 & 43.1 & 1.06 & 1.91 & 2.78 & $-$31.4\,\%\\
2,554 & 4 & 25 & 0.97 & 2.602 & 10.3$^\dagger$ & 15.5$^\dagger$ & 0.68 / 0.49 & 46.5 & 1.06 & 1.86 & 2.55 & $-$27.3\,\%\\
2,649 & 4 & 30 & 0.79 & 2.731 & 8.2$^\dagger$ & 43.6$^\dagger$ & 0.62 / 0.44 & 46.9 & 1.06 & 1.79 & 2.31 & $-$22.4\,\%\\
2,925 & 3 & 34 & 0.68 & 3.180 & 6.9$^\dagger$ & 40.0$^\dagger$ & 0.56 / 0.41 & 51.7 & 0.64 & 1.69 & 1.98 & $-$14.4\,\%\\
2,977 & 3 & 36 & 0.73 & 3.250 & 7.6$^\dagger$ & 37.2$^\dagger$ & 0.57 / 0.42 & 52.6 & 0.64 & 1.68 & 1.94 & $-$13.5\,\%\\
2,994 & 3 & 34 & 0.87 & 3.273 & 9.1$^\dagger$ & 14.4$^\dagger$ & 0.62 / 0.46 & 54.2 & 0.64 & 1.73 & 1.82 & $-$4.6\,\%\\
2,998 & 3 & 37 & 1.08 & 3.278 & 12.7$^\dagger$ & 33.4$^\dagger$ & 0.59 / 0.44 & 53.0 & 0.64 & 1.67 & 1.82 & $-$8.1\,\%\\
\bottomrule
\end{tabular}
}
\caption{Hardware projections for all 42 circuits. The reference row uses
this paper's reconstruction of the fastest evaluated serial window on the
surface code (3.10\,d, versus about 4.63\,d in~\cite{gidney2025}) and each
neutral-atom processor variant (separate from Cain et al.'s parallel
97-day design~\cite{cain2026}), and the published serial chip on LDPC-cat.
A dagger marks a neutral-atom extrapolation requiring an enlarged memory
($>1{,}480$ logical qubits). The paired LDPC-cat times retain the serial
residual overhead $H_0$ or set it to zero, and ``Cat ref.'' gives the
fastest listed single-chip configuration in Ref.~\cite{pappalardo2026}
fitting that physical footprint. ``Pin.\ ref.'' gives the fastest
Pinnacle~1.0~\cite{pinnacle2026} reference within that physical-qubit
budget.}
\label{tab:hw}
\end{table}

\subsection{Comparison at a fixed physical footprint}
The two LDPC-cat times in Table~\ref{tab:hw} use the same arithmetic with
different allowances for movement, preparation and measurement, and
Table~\ref{tab:hw} also lists the fastest published single-chip
configuration that fits each candidate's physical footprint. Retaining the
serial chip's residual overhead gives a shorter runtime than the reference
for 38 of the 42 circuits, whereas setting that overhead to zero changes
the ordering at three more widths (1,291 and 1,298 logical qubits at
$s=14$, and 1,796 at $s=8$; at 1,271 logical qubits the reference is faster
under either assumption). On Pinnacle, the candidate has a shorter projected
time at all 41 budgets with a feasible reference, with margins ranging from
4.6\,\% to 59.6\,\%.

%% file: tikz_phase.tex
% Appendix figure: coherent versus measurement-based erasure of a one-bit lookup.
\begin{tikzpicture}[
  font=\small, x=1cm, y=0.6cm,
  wire/.style={black!80, line width=0.55pt},
  gate/.style={draw=black!85, fill=white, line width=0.6pt, minimum height=4.4mm, minimum width=5mm, inner sep=1.5pt, font=\footnotesize},
  ctrl/.style={circle, fill=black, inner sep=0pt, minimum size=3.6pt},
  targ/.style={circle, draw=black, line width=0.55pt, inner sep=0pt, minimum size=3.2mm},
  cls/.style={double distance=1.1pt, black!80, line width=0.4pt},
]
\newcommand{\toff}[3]{% x, control rows (comma list given as two args), target row
  \node[ctrl] at (#1,#2) {}; \node[ctrl] at (#1,#2-1) {};
  \draw[wire] (#1,#2) -- (#1,#3-0.27);
  \node[targ] at (#1,#3) {}; \draw[wire] (#1,#3-0.27) -- (#1,#3+0.27);}
% ---------------------------------------------------------------- (a)
\node[anchor=west, font=\normalsize\bfseries] at (-1.0,1.1) {(a)};
\node[anchor=west, font=\footnotesize] at (-0.32,1.1) {Coherent erasure};
\foreach \y/\l/\r in {0/{$a_0$}/{}, -1/{$a_1$}/{}, -2/{$\ket{0}_w$}/{$\ket{0}_w$}, -3/{$D$}/{}}{
  \node[anchor=east, font=\footnotesize] at (-0.05,\y) {\l};
  \node[anchor=west, font=\footnotesize] at (4.55,\y) {\r};
  \draw[wire] (0,\y) -- (4.5,\y);}
\toff{0.9}{0}{-2}
\node[ctrl] at (2.25,-2) {}; \draw[wire] (2.25,-2) -- (2.25,-2.7);
\node[gate] at (2.25,-3) {$V$};
\toff{3.6}{0}{-2}
% ---------------------------------------------------------------- (b)
\begin{scope}[xshift=7.3cm]
\node[anchor=west, font=\normalsize\bfseries] at (-1.0,1.1) {(b)};
\node[anchor=west, font=\footnotesize] at (-0.32,1.1) {Measurement-assisted erasure};
\foreach \y/\l in {0/{$a_0$}, -1/{$a_1$}, -3/{$D$}}{
  \node[anchor=east, font=\footnotesize] at (-0.05,\y) {\l};
  \draw[wire] (0,\y) -- (6.2,\y);}
\node[anchor=east, font=\footnotesize] at (-0.05,-2) {$\ket{0}_w$};
\draw[wire] (0,-2) -- (3.75,-2);
\toff{0.9}{0}{-2}
\node[ctrl] at (2.0,-2) {}; \draw[wire] (2.0,-2) -- (2.0,-2.7);
\node[gate] at (2.0,-3) {$V$};
\node[gate] at (3.0,-2) {$H$};
\node[gate, minimum width=6.5mm] (m) at (4.1,-2) {};
\draw[line width=0.5pt] ([xshift=-2mm,yshift=-1mm]m.center) arc (160:20:2.1mm);
\draw[->, line width=0.5pt] ([yshift=-1.2mm]m.center) -- ++(55:2.4mm);
\draw[cls] (m.east) -- (5.35,-2) -- (5.35,-1.27);
\node[font=\scriptsize] at (4.95,-1.68) {$\mu$};
\node[ctrl] at (5.35,0) {}; \draw[wire] (5.35,0) -- (5.35,-0.75);
\node[gate, minimum width=5mm] at (5.35,-1) {$Z$};
\end{scope}
\end{tikzpicture}

%% file: ablation_table.tex
% Generated by scripts/ablation_summary.py from existing scores.
\begin{table}[htbp]
\centering\small
\begin{tabular}{rrrrrr}
\toprule
$w_1$ & Sharing $k$ & Checkpoints & Lazy & $Q$ & $T_{\rm fact}/10^9$\\
\midrule
6 & 1 & 0 & no & 1{,}432 & 5.688\\
6 & 1 & 4 & no & 1{,}520 & 5.055\\
6 & 1 & 4 & yes & 1{,}530 & 4.856\\
6 & 3 & 4 & no & 1{,}578 & 3.502\\
\midrule
8 & 6 & 4 & no & 1{,}690 & 2.756\\
8 & 6 & 4 & yes & 1{,}690 & 2.605\\
\bottomrule
\end{tabular}
\caption{Selected matched ablations from the scored search. All rows use
trimmed discrete-log registers, $s=8$, $\ell=21$, $w_{3a}=w_{3b}=3$,
$w_4=5$ and $f=33$. Compare rows within each $w_1$ block: $k=1$ means
no sharing. Lazy arithmetic requires all (four) checkpoints, so these are
selected conditional contrasts, not a full factorial design.}
\label{tab:ablations}
\end{table}